\documentclass[aps,prb,superscriptaddress,showpacs,floatfix,twocolumn]{revtex4-1}
\usepackage{graphicx,amsmath,amssymb,xspace,epsfig,float,multirow,subfigure,tabularx}

\pdfoutput=1

\begin{document}

\title{Triplet superconductivity in 3D Dirac semimetal due to exchange
interaction. }
\author{Baruch Rosenstein}
\email{vortexbar@yahoo.com }
\affiliation{Electrophysics Department, National Chiao Tung University, Hsinchu 30050, \textit{Taiwan,
R. O. C}}
\affiliation{Physics Department, Ariel University, Ariel 40700,
Israel}
\author{B. Ya. Shapiro}
\email{shapib@mail.biu.ac.il}
\affiliation{Physics Department, Bar-Ilan University, 52900 Ramat-Gan,
Israel}
\author{Dingping Li}
\email{lidp@pku.edu.cn}
\affiliation{School of Physics, Peking University, Beijing
100871, \textit{China}}
\affiliation{Collaborative Innovation Center of
Quantum Matter, Beijing, China}
\author{I. Shapiro}
\email{yairaliza@gmail.com}
\affiliation{Physics Department, Bar-Ilan University, 52900 Ramat-Gan,
Israel}

\begin{abstract}
Conventional phonon-electron interaction induces either triplet or one of
two (degenerate) singlet pairing states in time reversal and inversion
invariant 3D Dirac semi - metal. Investigation of the order parameters and
energies of these states at zero temperature in wide range of values of
chemical potential $\mu $, the effective electron-electron coupling constant
$\lambda $ and Debye energy $T_{D}$ demonstrates that when the exchange
interaction is neglected the singlet always prevails, however in significant
portions of the $\left( \mu ,\lambda ,T_{D}\right) $ parameter space the
energy difference is very small. This means that interactions that are small
but discriminate between the spin singlet and the spin triplet are important
in order to determine the nature of the superconducting order there. The
best candidate for such an interaction in materials under consideration is
the exchange (the Stoner term) characterized by constant $\lambda _{ex}$. We
show that at values of $\lambda _{ex}$ much smaller then ones creating
Stoner instability to ferromagnetism $\lambda _{ex}\sim 1$ the triplet
pairing becomes energetically favored over the singlet ones for $\mu <T_{D}$%
. The 3D quantum critical point at $\mu =0$ is considered in detail.
\end{abstract}

\maketitle

\section{Introduction.}

Recently solids with electronic states described by the Bloch wave
functions, obeying the "pseudo-relativistic" Dirac equation (with Fermi
velocity $v_{F}$ replacing the velocity of light) attracted widespread
attention. One outstanding example is graphene, a two-dimensional (2D)
hexagonal lattice made of carbon atoms. The effective low energy two-band
model (near its $K$ and $K^{\prime }$ points in the Brillouin zone) is
described by the four-component (two pseudospins/sublattices and two
valleys) massless 2D Dirac Hamiltonian (in fact there are two species of
such quasiparticles for each spin). Although a similar two-band electronic
structure of bismuth was described by a four-component nearly massless Dirac
fermion in 3D caused by spin-orbit interaction long ago \cite{Wolff} (with
spin replacing pseudospin), only recently several systems were
experimentally found to exhibit the 3D Dirac quasiparticles. Their discovery
followed recent exploration of the topological band theory\cite{Zhang}.

One of the effective ideas to get a 3D Dirac semi-metal is to close the
insulating gap by tuning a topological insulator towards the quantum phase
transition to trivial insulators when the reflection symmetry is preserved%
\cite{Kane}. The time reversal invariant 3D Dirac point in materials like $%
Na_{3}Bi$ was theoretically investigated\cite{Wang13} and experimentally
observed\cite{Liu}. A well known compound $Cd_{3}As_{2}$ is a
symmetry-protected 3D Dirac semimetal with a single pair of Dirac points in
the bulk and nontrivial Fermi arcs on the surface \cite{Fang}. Most recently
conductivity and magneto-absorption of a zinc-blende crystal, $HgCdTe,$ was
measured\cite{Potemski} and is in agreement with theoretical expectations in
Dirac semimetal \cite{Wan}. Ab-initio calculations and symmetry arguments
predict\cite{Kane2} that cristobalite $BiO_{2}$ exhibits Dirac points at
three symmetry related $X$ points on the boundary of the FCC Brillouin zone.
Pyrochlore iridates \cite{Viswanath} and inverse perovskites \cite{Ogata}
were also predicted to be Weyl - semimetals. Several known materials with
well known magnetic or transport properties have recently undergone a
"delayed" realization that they are actually Dirac semimetals\cite{quasi}.

The discovery of the 3D Dirac materials makes it possible to study their
physics including remarkable electronic properties. This is rich in new
phenomena, not seen in 2D Dirac semi - metals like graphene and surface
states of topological insulator also harboring 2D Weyl quasiparticles.
Examples include the giant diamagnetism that diverges logarithmically when
the chemical potential approaches the 3D Dirac point; linear-in-frequency AC
conductivity that has an imaginary part\cite{Wan}; quantum magnetoresistance
showing linear field dependence in the bulk\cite{Ogata}. Most of the
properties of these new materials were measured at relatively high
temperatures. However some of the topological insulators and suspected 3D
Dirac semi-metals exhibit superconductivity at about the liquid $He$
temperature.

The well known topological insulator $Bi_{2}Se_{3}$ doped with $Cu$, becomes
superconducting at $T_{c}=3.8K$\cite{Ong}. At present its pairing symmetry
is unknown. Some experimental evidence\cite{phononexp} point to a
conventional phononic pairing mechanism. The spin independent part of the
effective electron - electron interaction due to phonons was studied
theoretically\cite{DasSarma13,Guinea}.\ For a conventional parabolic
dispersion relation, typically independent of spin, the phonon mechanism
leads to the $s$-wave superconductivity. The layered, non-centrosymmetric
heavy element $PbTaSe_{2}$ was found to be superconducting \cite{Cava}. Its
electronic properties like specific heat, electrical resistivity, and
magnetic-susceptibility indicate that $PbTaSe_{2}$ is a moderately coupled,
type-II BCS superconductor with large electron-phonon coupling constant of $%
\lambda =0.74$. It was shown theoretically to possess a very asymmetric 3D
Dirac point created by strong spin-orbit coupling. If the 3D is confirmed,
it might indicate that the superconductivity is conventional phonon mediated.

More recently when the $Cu$ doped $Bi_{2}Se_{3}$ was subjected to pressure%
\cite{pressureBiSe}, $T_{c}$ increased to $7K$ at $30GPa$. Quasilinear
temperature dependence of the upper critical field $H_{c2},$ exceeding the
orbital and Pauli limits for the singlet pairing, points to the triplet
superconductivity. The band structure of the superconducting compounds is
apparently not very different from its parent compound $Bi_{2}Se_{3}$, so
that one can keep the two band $\mathbf{k}\cdot \mathbf{p}$ description ($Se$
$p_{z}$ orbitals on the top and bottom layer of the unit cell mixed with its
neighboring $Bi$ $p_{z}$ orbital). Electronic-structure calculations of the
compound under pressure\cite{pressureBiSe} reveal a single bulk
three-dimensional Dirac cone like in $Bi$ with large spin-orbit coupling.
Usually the phonon mediated pairing leads to the $s$-wave "conventional"
superconductivity, while the $p$-wave pairing in $SrRuO_{3}$ or heavy
fermion superconductors like $UPt_{3}$ "unconventional" nonphononic
mechanism typically hinges on nonlocal interactions.

The case of the Dirac semi-metals is very special due to the strong spin
dependence of the itinerant electrons' effective Hamiltonian. It was pointed
out\cite{FuBerg,Herbut} that in this case the triplet possibility can arise
although the triplet gap is smaller than that of the singlet, the difference
sometimes is not large for spin independent electron - electron
interactions. Very recently the spin dependent part of the phonon induced
electron - electron interaction was considered\cite{DasSarma14} and it was
shown that the singlet is still favored over the triplet pairing. Another
essential spin dependent effective electron-electron interaction is the
Stoner exchange among itinerant electrons\cite{White} leading to
ferromagnetism in transition metals. While in the best 3D Weyl semimetal
candidates it is too small to form a ferromagnetic state, it might be
important to determine the nature of the superconducting condensate.
Obviously it favors the triplet pairing.

It therefore of importance to clarify theoretically two questions. (i) Does
a conventional phononic superconductivity exist in these materials with just
a minute density of states compared even with high $T_{c}$ cuprates that
apparently utilize much more powerful pairing mechanism than phonons offer?
(ii) Is it possible that phonons in 3D Dirac materials lead to triplet
pairing that even becomes dominant under certain circumstances?

In the present paper we construct the theory of the superconducting
transition in 3D Dirac semi-metal at arbitrary chemical potential including
zero, assuming the local (probably, but not necessarily, phonon mediated)
pairing. The possible pairing channels are classified in this rather unusual
situation using symmetries of the system. In contrast to the 2D case, the
triplet pairing is not only possible, but for a moderately strong exchange
interaction is the preferred channel taking over the more "conventional"
singlet one.

It turns out that the triplet superconductivity is easier realized in the
intriguing case of a small chemical potential. The superconductivity there
is governed by{\LARGE \ }a quantum critical point (QCP)\cite{Sachdev}. The
concept of QCP at zero temperature and varying doping constitutes a very
useful language for describing the microscopic origin of superconductivity
in high $T_{c}$ cuprates and other "unconventional" superconductors\cite{Wen}%
. Quantum criticality, although occurring often in 2D (even in the context
of surface superconductivity in topological insulators\cite{Li14}), is very
rare in 3D. We find and characterize the quantum critical points
corresponding to both the singlet and the triplet superconducting
transitions. There are experimental methods to tune the chemical potential
by doping (for example by copper\cite{Ong}), gating\cite{bias}, pressure\cite%
{pressureBiSe} etc.

The paper is organized as follows. The model of the phonon mediated and
exchange effective local interactions of Dirac fermion is presented and the
method of its solution (in the Gorkov equations form) including the symmetry
analysis of possible pairing channels is given in Section II. In Section III
the phase diagram for spin independent interactions is established and the
regions in parameter space where singlet and triplet states are nearly
degenerate are identified. The Stoner exchange interaction is considered
perturbatively in Section IV. A novel case of zero chemical potential (QCP)
is studied in Section V. Section VI contains discussion on experimental
feasibility of the phonon mediated surface superconductivity in 3D Weyl
semi-metals, as well as a comparison with earlier calculations and\
conclusion.

\section{The local pairing model in the Dirac semimetal.}

\subsection{Interactions in the Dirac semi-metal.}

Electrons in the 3D Dirac semimetal are described by fields operators $\psi
_{fs}\left( \mathbf{r}\right) $, where $f=L,R$ are the valley index
(pseudospin) for the left/right chirality bands with spin projections taking
the values $s=\uparrow ,\downarrow $ with respect to, for example, $z$ axis.
To use the Dirac ("pseudo-relativistic") notations, these are combined into
a four component bi-spinor creation operator, $\psi ^{\dagger }=\left( \psi
_{L\uparrow }^{\dagger },\psi _{L\downarrow }^{\dagger },\psi _{R\uparrow
}^{\dagger },\psi _{R\downarrow }^{\dagger }\right) $, whose index $\gamma
=\left\{ f,s\right\} $ takes four values. The non-interacting massless
Hamiltonian with Fermi velocity $v_{F}$ and chemical potential $\mu $ reads%
\cite{Wang13},%
\begin{equation}
K=\int_{\mathbf{r}}\psi ^{+}\left( \mathbf{r}\right) \widehat{K}\psi \left(
\mathbf{r}\right) \text{;\ \ \ \ \ \ }\widehat{K}_{\gamma \delta }=-i\hbar
v_{F}\nabla ^{i}\alpha _{\gamma \delta }^{i}-\mu \delta _{\gamma \delta }%
\text{,}  \label{kinetic}
\end{equation}%
where three $4\times 4$ matrices, $i=x,y,z$,
\begin{equation}
\mathbf{\alpha }=\left(
\begin{array}{cc}
\mathbf{\sigma } & 0 \\
0 & -\mathbf{\sigma }%
\end{array}%
\right) \text{,}  \label{alpha}
\end{equation}%
are presented in the block form via Pauli matrices $\mathbf{\sigma }$. They
are related to the Dirac $\mathbf{\gamma }$ matrices (in the chiral
representation, sometimes termed "spinor") by $\ \mathbf{\alpha }=\beta
\mathbf{\gamma }$ with%
\begin{equation}
\beta =\left(
\begin{array}{cc}
0 & \mathbf{1} \\
\mathbf{1} & 0%
\end{array}%
\right) \text{.}  \label{beta}
\end{equation}%
Here $\mathbf{1}$ is $2\times 2$ identity matrix.

We consider a special case of 3D rotational symmetry that in particular has
an isotropic Fermi velocity. Moreover we assume the time reversal, $\Theta
\psi \left( \mathbf{r}\right) =i\sigma _{y}\psi ^{\ast }\left( \mathbf{r}%
\right) $, and inversion symmetries although the pseudo Lorentz symmetry
will be explicitly broken by interactions. The spectrum of single particle
excitations is linear, see Fig.1. The chemical potential $\mu $ is counted
from the Dirac point.

Electrons interact electrostatically via the density - density potential $%
v\left( \mathbf{r}\right) $:%
\begin{eqnarray}
V_{e-e} &=&\frac{1}{2}\int_{\mathbf{rr}^{\prime }}\rho \left( \mathbf{r}%
\right) v\left( \mathbf{r-r}^{\prime }\right) \rho \left( \mathbf{r}^{\prime
}\right) ;  \label{ee} \\
\rho \left( \mathbf{r}\right) &=&\psi _{\alpha }^{+}\left( \mathbf{r}\right)
\psi _{\alpha }\left( \mathbf{r}\right) =\psi _{Ls}^{+}\psi _{Ls}+\psi
_{Rs}^{+}\psi _{Rs}  \notag
\end{eqnarray}%
In Weyl semi-metals there is no static screening at $\mu =0$, although
dynamically it is screened within RPA \cite{Viswanath}. The screening length
is therefore not small like in good metals. However in most materials that
realize the Dirac semi - metals \cite{Ogata}, there is a large dielectric
constant $\varepsilon \sim 50$ that allows phonon - electron coupling \cite%
{AGD},%
\begin{equation}
H_{e-ph}=w\int \rho \left( \mathbf{r}\right) \mathbf{\nabla }\cdot \mathbf{u}%
\left( \mathbf{r}\right) \text{,}  \label{eph}
\end{equation}%
to overpower it to create the Cooper pairs as mentioned in Introduction.
Here $\mathbf{u}\left( \mathbf{r}\right) $ denotes the displacement of ions
and the electron-ion coupling $w\propto M^{-1/2}$, where $M$ is the ion
mass. The effective electron-electron interaction due to both electron -
phonon attraction and Coulomb repulsion (pseudopotential) can be generally
expanded in derivatives. The leading term usually called the local (or the $%
s $-wave pairing) coupling is

\begin{eqnarray}
V_{eff} &=&-\frac{g}{2}\int_{\mathbf{rr}^{\prime }}\rho \left( \mathbf{r}%
\right) \delta \left( \mathbf{r-r}^{\prime }\right) \rho \left( \mathbf{r}%
^{\prime }\right) =  \label{local} \\
&=&\mathbf{-}\frac{g}{2}\int_{\mathbf{r}}\psi _{\alpha }^{+}\left( \mathbf{r}%
\right) \psi _{\beta }^{+}\left( \mathbf{r}\right) \psi _{\beta }\left(
\mathbf{r}\right) \psi _{\alpha }\left( \mathbf{r}\right) \text{.}
\end{eqnarray}%
Unlike the free Hamiltonian $K$, Eq.(\ref{kinetic}), this interaction
Hamiltonian does not mix different spin components.

Such a coupling implicitly restricts the spin independent local interaction
to be symmetric under the band permutation (the constants in front of the
interband $\psi _{1}^{\dagger }\psi _{1}\psi _{2}^{\dagger }\psi _{2}$ and
intraband $\psi _{1}^{\dagger }\psi _{1}\psi _{1}^{\dagger }\psi _{1}$ terms
are the same). If the mechanism of pairing is due to acoustic phonons only,
such an additional term is not generated. A more general case with
additional independent term was considered in ref.\cite{FuBerg}.

Usually such coupling with a positive coupling constant $g$ leads to the $s$%
-wave "conventional" pairing, while an "unconventional" $p$-wave pairing in $%
SrRuO_{3}$ or heavy fermion superconductors like $UPt_{3}$ requires
subleading interaction terms with two derivatives (most probably beyond
electron - phonon mechanism). Two qualitatively different cases will be
considered, see Fig.1. When the chemical potential $\mu $ is much larger
than the Debye energy $T_{D}$ characterizing the outreach of the phonon -
electron coupling, see Fig.1a, the physics is similar to that considered for
the parabolic bands within the BCS approximations \cite{AGD}. The opposite
case , $\mu <T_{D}$ (Fig. 1b) is unusual and most of our findings are
devoted to this case.

The Coulomb forces in Eq.(\ref{ee}) in addition to direct repulsion lead to
spin dependent forces due to exchange. The exchange interaction among
itinerant electrons first considered by Stoner \cite{White}, although small
and unable to form a ferromagnetic state in materials under consideration,
will be important for the nature of the condensate since it will lift the
degeneracy between the singlet and the triplet pairing:%
\begin{equation}
V_{s-s}=-\frac{1}{2}\int_{\mathbf{r,r}^{\prime }}J\left( \mathbf{r-r}%
^{\prime }\right) \mathbf{S}\left( \mathbf{r}\right) \cdot \mathbf{S}\left(
\mathbf{r}^{\prime }\right) \text{.}  \label{s-s}
\end{equation}%
Spin density in Dirac semi-metal has the form
\begin{equation}
\mathbf{S}\left( \mathbf{r}\right) =\psi ^{+}\left( \mathbf{r}\right)
\mathbf{\Sigma }\psi \left( \mathbf{r}\right) \text{,}  \label{spindensity}
\end{equation}%
where the matrices
\begin{equation}
\mathbf{\Sigma =-\alpha }\gamma _{5}=\left(
\begin{array}{cc}
\mathbf{\sigma } & 0 \\
0 & \mathbf{\sigma }%
\end{array}%
\right) ;\text{ \ \ \ }\gamma _{5}=\left(
\begin{array}{cc}
\mathbf{-}1 & 0 \\
0 & 1%
\end{array}%
\right) ,  \label{sigma_matrices}
\end{equation}%
are also the rotation generators.
\begin{figure}[tbp]
\centering
\subfigure[]{\includegraphics[width=6cm]{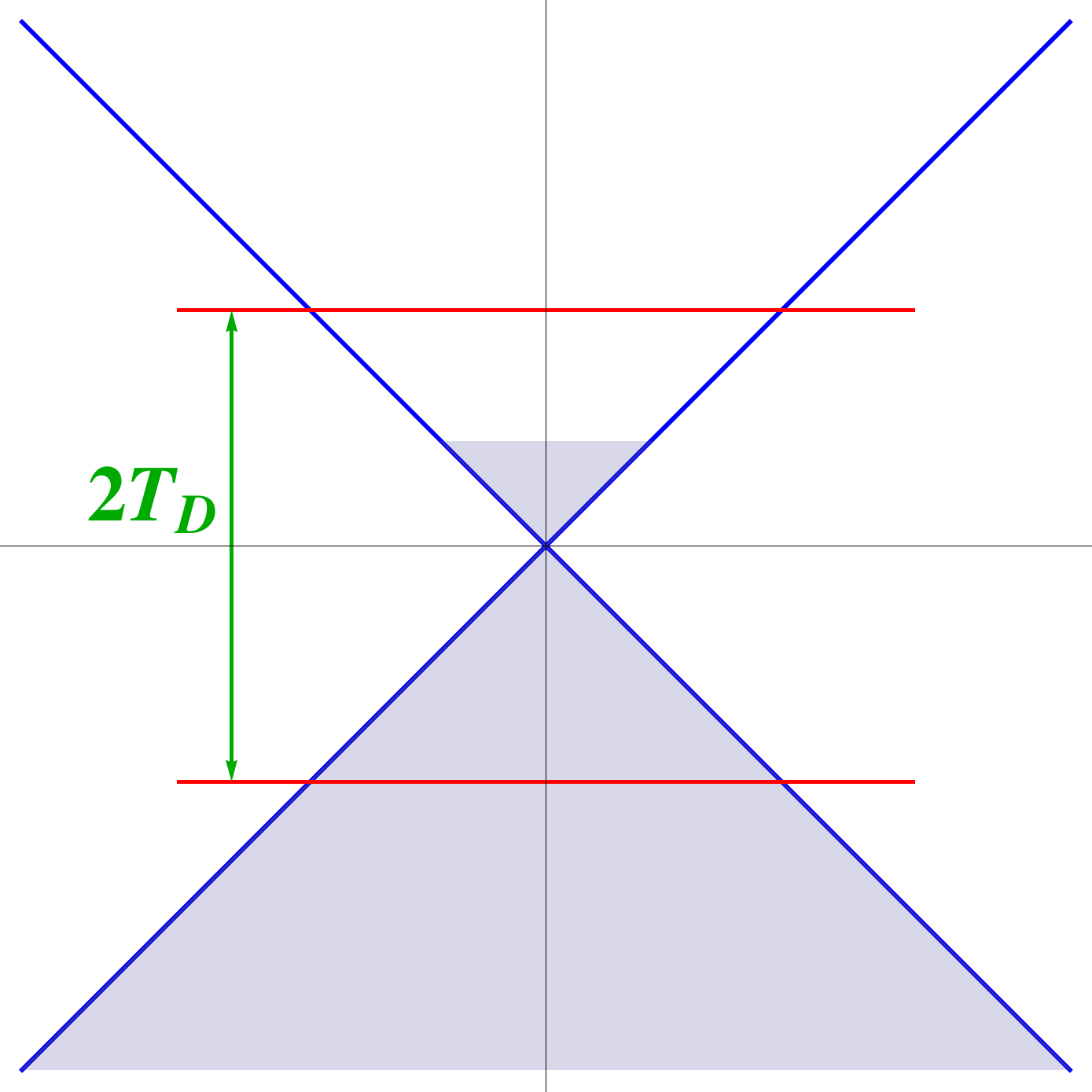}} \subfigure[]{%
\includegraphics[width=6cm]{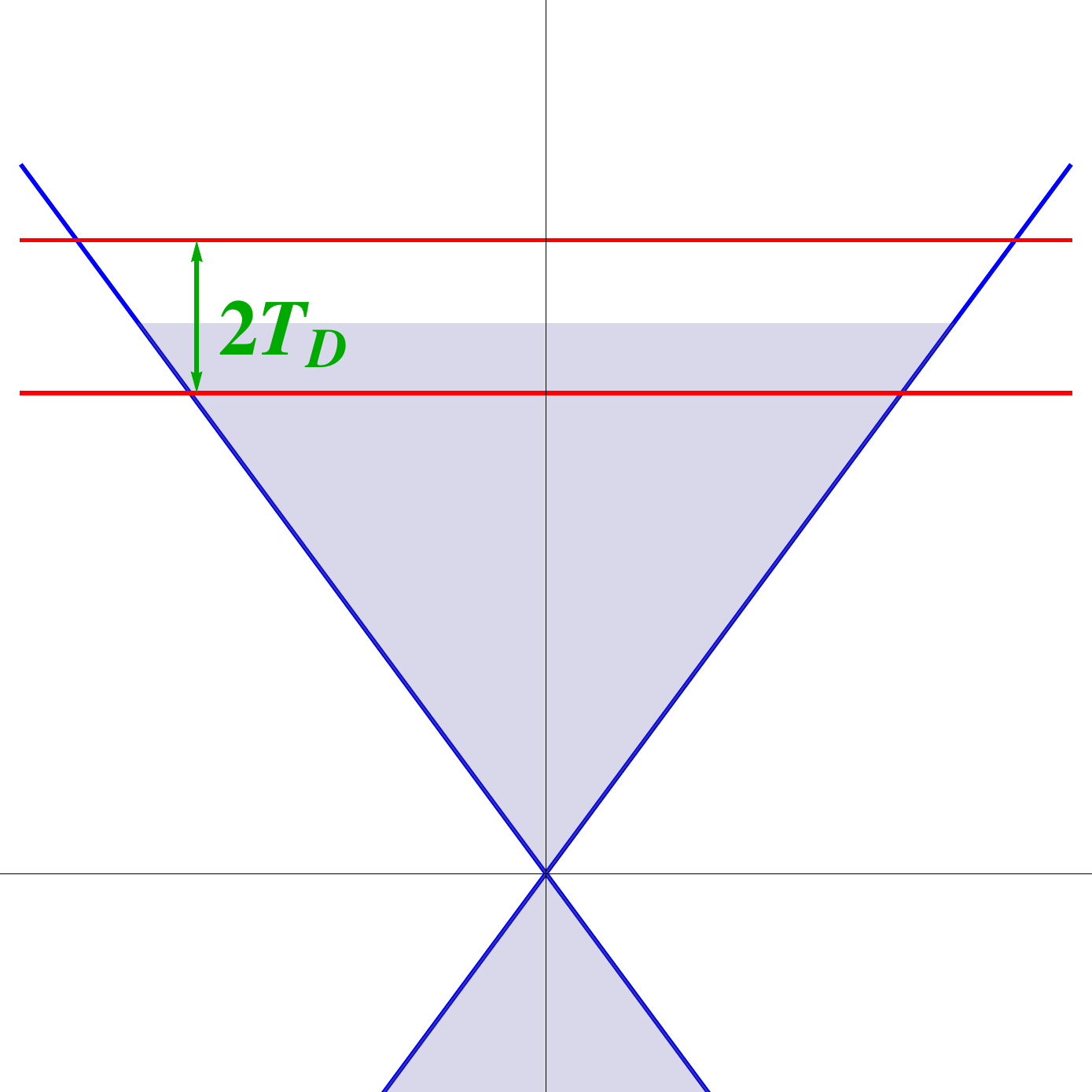}}
\caption{Chemical potential in Dirac semi - metals and the phonon mediated
pairing. (a) Chemical potential relative to Dirac point is smaller that
typical energy of phonons, the Debye energy $T_{D}$. (b) The BCS
approximation limit: the chemical potential is much larger than the Debye
energy $T_{D}$.}
\end{figure}

\subsection{The symmetry classification of possible pairing channels.}

Since we consider the local interactions as dominant, the superconducting
condensate (the off-diagonal order parameter) will be local%
\begin{equation}
O=\int_{\mathbf{r}}\psi _{\alpha }^{+}\left( \mathbf{r}\right) M_{\alpha
\beta }\psi _{\beta }^{+}\left( \mathbf{r}\right) ,  \label{O}
\end{equation}%
where the constant matrix $M$ should be a $4\times 4$ antisymmetric matrix.
Due to the rotation symmetry they transform covariantly under infinitesimal
rotations generated by the spin $S^{i}$ operator, Eq.(\ref{spindensity}) :

\begin{eqnarray}
&&\int_{\mathbf{r,r}^{\prime }}\left[ \psi _{\alpha }^{+}\left( r\right)
M_{\alpha \beta }\psi _{\beta }^{+}\left( r\right) ,\psi _{\gamma
}^{+}\left( r^{\prime }\right) \mathbf{\Sigma }_{\gamma \delta }^{i}\psi
_{\delta }\left( r^{\prime }\right) \right]  \label{transformation} \\
&=&\int_{r}\psi _{\gamma }^{+}\left( r\right) \mathbf{\Sigma }_{\gamma
\delta }^{i}\left\{ M_{\delta \kappa }^{t}-M_{\delta \kappa }\right\} \psi
_{\kappa }^{+}\left( r\right) \text{.}
\end{eqnarray}%
The representations of the rotation group therefore characterize various
possible superconducting phases.

Out of 16 matrices of the four dimensional Clifford algebra six are
antisymmetric and one finds one vector and three scalar multiplets of the
rotation group. The multiplets contain:

(i) a triplet of order parameters:
\begin{equation}
\left\{ M_{x}^{T},M_{y}^{T},M_{z}^{T}\right\} =\left\{ \beta \alpha
_{z},-i\beta \gamma _{5},\beta \alpha _{x}\right\}  \label{triplet}
\end{equation}%
The algebra is
\begin{equation}
\left[ M_{i}^{T},\Delta _{j}^{T}\right] =i\varepsilon _{ijk}M_{k}^{T}\text{.}
\label{algebra}
\end{equation}

(ii) three singlets

\begin{equation}
M_{1}^{S}=i\alpha _{y};\text{ \ \ }M_{2}^{S}=i\Sigma _{y};\text{ \ \ }%
M_{3}^{S}=-i\beta \alpha _{y}\gamma _{5}\text{.}  \label{singlets}
\end{equation}%
Which one of the condensates is realized at zero temperature is determined
by the parameters of the Hamiltonian and is addressed within the Gaussian
approximation next.

\section{The phase diagram for spin independent interactions.}

\subsection{Gorkov equations.}

The BCS type approximation can be employed. Using the standard formalism,
the Matsubara Green's functions ($\tau $ is the Matsubara time)
\begin{eqnarray}
G_{\alpha \beta }\left( \mathbf{r},\tau ;\mathbf{r}^{\prime },\tau ^{\prime
}\right)  &=&-\left\langle T_{\tau }\psi _{\alpha }\left( \mathbf{r},\tau
\right) \psi _{\beta }^{\dagger }\left( \mathbf{r}^{\prime },\tau ^{\prime
}\right) \right\rangle \text{;}  \label{GFdef} \\
F_{\alpha \beta }^{\dagger }\left( \mathbf{r},\tau ;\mathbf{r}^{\prime
},\tau ^{\prime }\right)  &=&\left\langle T_{\tau }\psi _{\alpha }^{\dagger
}\left( \mathbf{r},\tau \right) \psi _{\beta }^{\dagger }\left( \mathbf{r}%
^{\prime },\tau ^{\prime }\right) \right\rangle \text{,}  \notag
\end{eqnarray}%
obey the Gor'kov equations\cite{AGD}:%
\begin{gather}
-\frac{\partial G_{\gamma \kappa }\left( \mathbf{r},\tau ;\mathbf{r}^{\prime
},\tau ^{\prime }\right) }{\partial \tau }-\int_{\mathbf{r}^{\prime \prime
}}\left\langle \mathbf{r}\left\vert \widehat{K}_{\gamma \beta }\right\vert
\mathbf{r}^{\prime \prime }\right\rangle G_{\beta \kappa }\left( \mathbf{r}%
^{\prime \prime },\tau ;\mathbf{r}^{\prime },\tau ^{\prime }\right)   \notag
\\
-gF_{\beta \gamma }\left( \mathbf{r},\tau ;\mathbf{r},\tau \right) F_{\beta
\kappa }^{\dagger }\left( \mathbf{r},\tau ,\mathbf{r}^{\prime },\tau
^{\prime }\right) =\delta ^{\gamma \kappa }\delta \left( \mathbf{r-r}%
^{\prime }\right) \delta \left( \tau -\tau ^{\prime }\right) ;  \notag \\
\frac{\partial F_{\gamma \kappa }^{\dagger }\left( \mathbf{r},\tau ;\mathbf{r%
}^{\prime },\tau ^{\prime }\right) }{\partial \tau }-\int_{\mathbf{r}%
^{\prime \prime }}\left\langle \mathbf{r}\left\vert \widehat{K}_{\gamma
\beta }^{t}\right\vert \mathbf{r}^{\prime \prime }\right\rangle F_{\beta
\kappa }^{\dagger }\left( \mathbf{r}^{\prime \prime },\tau ;\mathbf{r}%
^{\prime },\tau ^{\prime }\right)   \notag \\
-gF_{\gamma \beta }^{\dagger }\left( \mathbf{r},\tau ;\mathbf{r},\tau
\right) G_{\beta \kappa }\left( \mathbf{r},\tau ,\mathbf{r}^{\prime },\tau
^{\prime }\right) =0\text{.}  \label{Gorkov}
\end{gather}%
In the homogeneous case the Gor'kov equations for Fourier components of the
Greens functions simplify considerably,
\begin{eqnarray}
D_{\gamma \beta }^{-1}G_{\beta \kappa }\left( \omega ,p\right) -\Delta
_{\gamma \beta }F_{\beta \kappa }^{\dagger }\left( \omega ,p\right)
&=&\delta ^{\gamma \kappa }\text{;}  \label{Gorkov_uniform} \\
D_{\beta \gamma }^{-1}F_{\beta \kappa }^{\dagger }\left( \omega ,p\right)
+\Delta _{\gamma \beta }^{\ast }G_{\beta \kappa }\left( \omega ,p\right)
&=&0\text{,}  \notag
\end{eqnarray}%
where $\omega =\pi T\left( 2n+1\right) $ is the Matsubara frequency and$\
D_{\gamma \beta }^{-1}=\left( i\omega -\mu \right) \delta _{\gamma \beta
}+v_{F}p^{j}\alpha _{\alpha \beta }^{j}$.

\bigskip The matrix gap function can be chosen as ($d$ is real)
\begin{equation}
\Delta _{\beta \gamma }=gF_{\gamma \beta }\left( 0\right) =gdM_{\gamma \beta
}\text{.}  \label{delta}
\end{equation}%
These equations are conveniently presented in matrix form (superscript $t$
denotes transposed and $I$ - the identity matrix):
\begin{eqnarray}
D^{-1}G-\Delta F^{\dagger } &=&I\text{;}  \label{matrixeq} \\
D^{t-1}F^{\dagger }+\Delta ^{\ast }G &=&0\text{.}  \notag
\end{eqnarray}%
Solving these equations one obtains
\begin{eqnarray}
G^{-1} &=&D^{-1}+\Delta D^{t}\Delta ^{\ast }\text{;}  \label{solution} \\
F^{\dagger } &=&-D^{t}\Delta ^{\ast }G\text{,}  \notag
\end{eqnarray}%
with the gap function found from the consistency condition
\begin{equation}
\Delta ^{\ast }=-g\sum\limits_{\omega q}D^{t}\Delta ^{\ast }G\text{.}
\label{gap eq}
\end{equation}%
Now we find solutions of this equation for each of the possible
superconducting phases.

\subsection{Triplet solution of gap equation.\textit{\ \ }}

In this phase rotational symmetry is spontaneously broken simultaneously
with the electric charge $U\left( 1\right) $ (global gauge invariance)
symmetry. Assuming $z$ direction of the $p$ - wave condensate the order
parameter matrix takes a form: $\Delta =\Delta _{T}M_{z}^{T}=\Delta
_{T}\beta \alpha _{x}$. In this Section we use the units of $v_{F}=1,\hbar
=1 $ and the energy scale will be set by the Debye cutoff, $T_{D}=1$, of the
electron - phonon interactions, see below. The off-diagonal ($41$) matrix
element of the matrix gap equation, Eq.(\ref{gap eq}), for real $\Delta
_{T}>0$ is:
\begin{gather}
\frac{1}{g}=\sum\limits_{\omega q}\left( \Delta _{T}^{2}+p_{\perp
}^{2}-p_{z}^{2}+\mu ^{2}+\omega ^{2}\right) \times \left[ \left( \Delta
_{T}^{2}+\omega ^{2}\right) ^{2}\right.  \label{gap} \\
+\left. \left( p^{2}-\mu ^{2}\right) ^{2}+2\left( p^{2}+\mu ^{2}\right)
\omega ^{2}+2\Delta _{T}^{2}\left( p_{\perp }^{2}-p_{z}^{2}+\mu ^{2}\right)
\right] ^{-1}\text{,}  \notag
\end{gather}%
where $p_{\perp }^{2}=p_{x}^{2}+p_{y}^{2}$. The spectrum of elementary
excitations obtained from the four poles of the Green's function, see Fig.2,
is\ (in physical units)
\begin{equation}
E_{\pm }^{2}=\Delta _{T}^{2}+v_{F}^{2}p^{2}+\mu ^{2}\pm 2v_{F}\sqrt{\Delta
_{T}^{2}p_{z}^{2}+p^{2}\mu ^{2}}\text{.}  \label{spectrum}
\end{equation}

\begin{figure}[tbp]
\centering
\subfigure[]{\includegraphics[width=6cm]{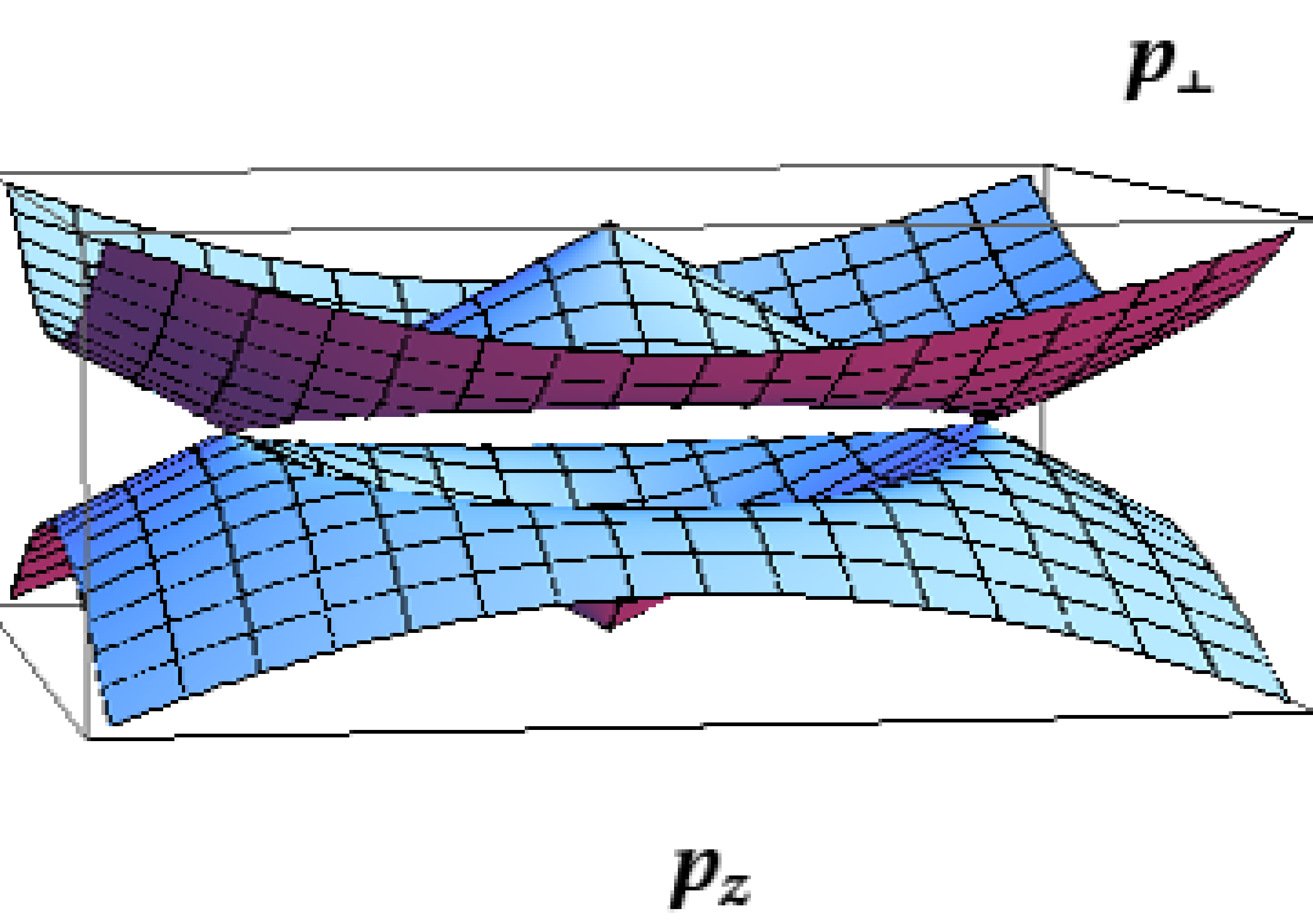}} \subfigure[]{%
\includegraphics[width=6cm]{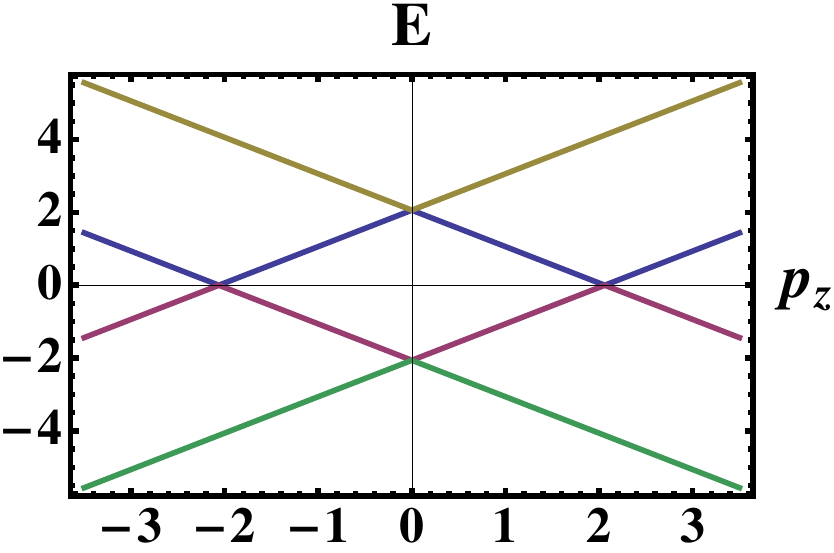}}
\caption{Spectrum of triplet excitations (a) section $p_{\perp }=0$. (b)
There is also a saddle points with energy gap. }
\end{figure}
There are two nodes at $p_{x}=p_{y}=0,v_{F}p_{z}=\pm \sqrt{\Delta
_{T}^{2}+\mu ^{2}}$, when the branches $+\left\vert E_{-}\right\vert $ and $%
-\left\vert E_{-}\right\vert $ cross, see Fig.2a and a section $p_{\perp }=0$
in Fig.2b. There is also a saddle point with energy gap, $2\Delta _{T}$ on
the circle $p_{x}^{2}+p_{y}^{2}=\mu ^{2},p_{z}=0,$ see the section in the $%
p_{z}=0$ direction in Fig. 3a. The higher energy band $E_{+}$ touches the
lower band at $p=0$, so that there is a Dirac point for quasiparticles, see
Fig. 3b.

\begin{figure}[tbp]
\centering
\subfigure[]{\includegraphics[width=6cm]{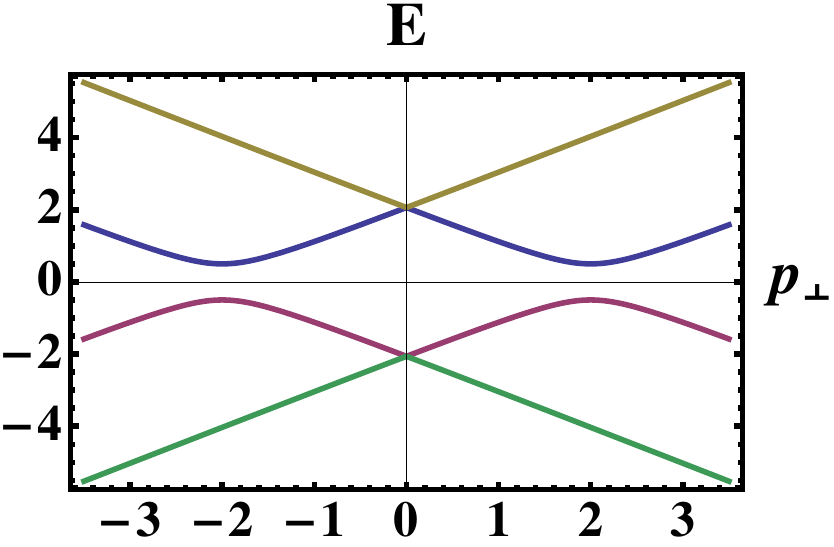}} \subfigure[]{%
\includegraphics[width=6cm]{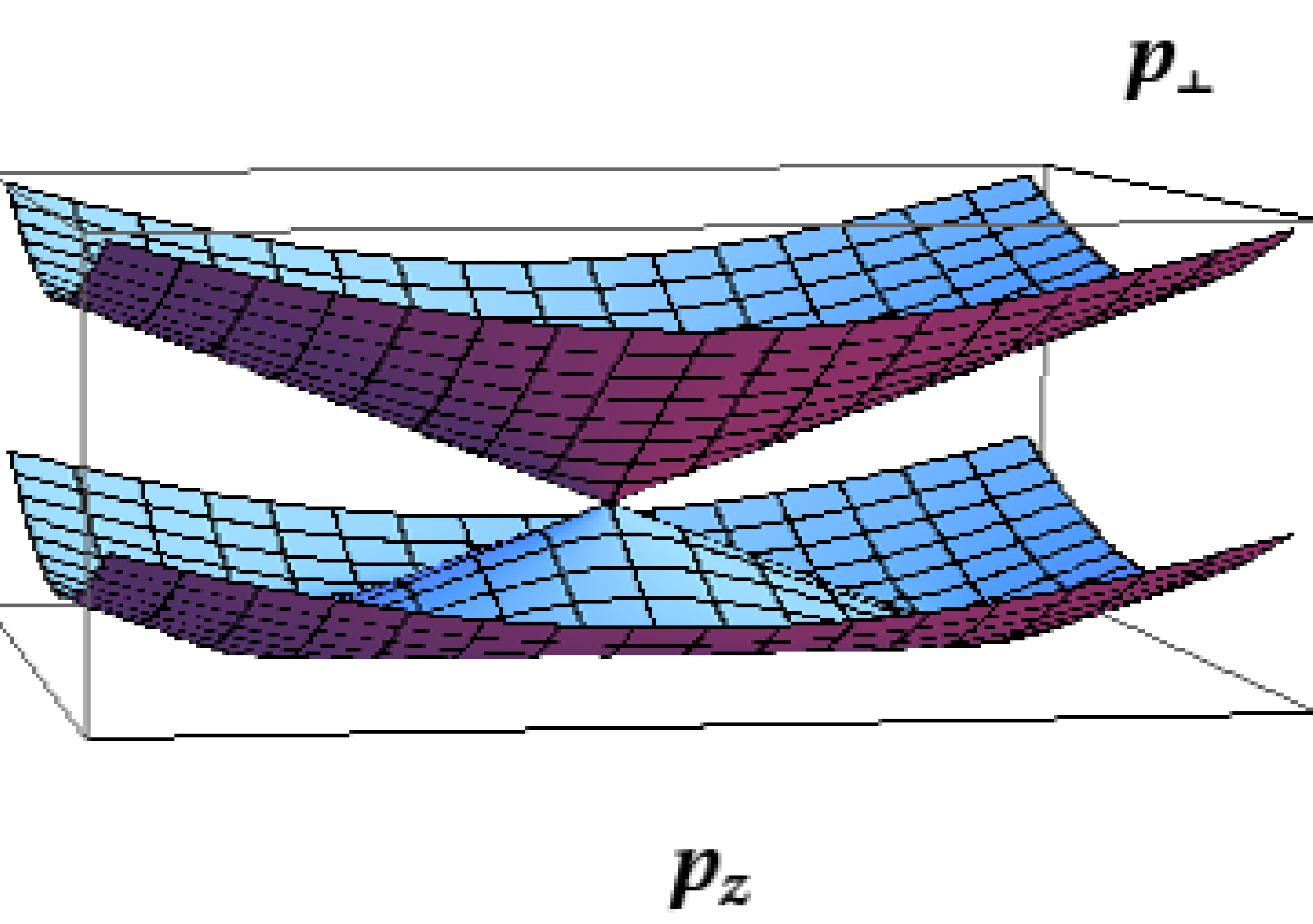}}
\caption{Spectrum of triplet excitations (a) $2\Delta _{T}$ on the circle $%
p_{x}^{2}+p_{y}^{2}=\protect\mu ^{2},p_{z}=0$ see the section in the $%
p_{z}=0 $ direction. (b) The higher energy band $E_{+}$ touches the lower
band at $p=0$, so that there is a Dirac point for quasiparticles.}
\end{figure}

Integration over $\omega ,$ using polar coordinates for $p$ and $x=\cos
\theta ,$ $\zeta =\sqrt{\Delta _{T}^{2}x^{2}+\mu ^{2}}$, gives%
\begin{gather}
\frac{1}{g}=\frac{1}{8\pi ^{2}}\int_{p=\max \left[ \mu -1,0\right] }^{\mu
+1}\int_{x=0}^{1}\frac{p^{2}}{\zeta }\left\{ \frac{\zeta +px^{2}}{\sqrt{%
\Delta _{T}^{2}+p^{2}+\mu ^{2}+2p\zeta }}\right.   \notag \\
+\left. \frac{\zeta -px^{2}}{\sqrt{\Delta _{T}^{2}+p^{2}+\mu ^{2}-2p\zeta }}%
\right\} \text{.}  \label{gapeqT}
\end{gather}%
The lower bound on the momentum integration is nonzero when the chemical
potential $\mu $ exceeds $T_{D}$. The integral over $x$ was performed
analytically, while the last integral was done numerically. The result of
the numerical solution of the gap equation for $\Delta _{T}$ is presented in
Fig.4a. The lines of fixed $g$ in the $\mu -\Delta _{T}$ plane are shown. As
expected the gap increases as a function of $\mu $. However, when the same
is replotted as lines of fixed phonon-electron coupling,
\begin{equation}
\lambda =gD\left( \mu \right) =g\mu ^{2}/\left( 8\pi ^{2}v_{F}^{3}\hbar
^{3}\right) ,  \label{lambda}
\end{equation}%
the gap increases upon reduction in $\mu $, see Fig.4b. At large $\mu >>T_{D}
$ the gap becomes independent of $\mu $ as in BCS, discussed next.
\begin{figure}[tbp]
\centering
\subfigure[]{\includegraphics[width=6cm]{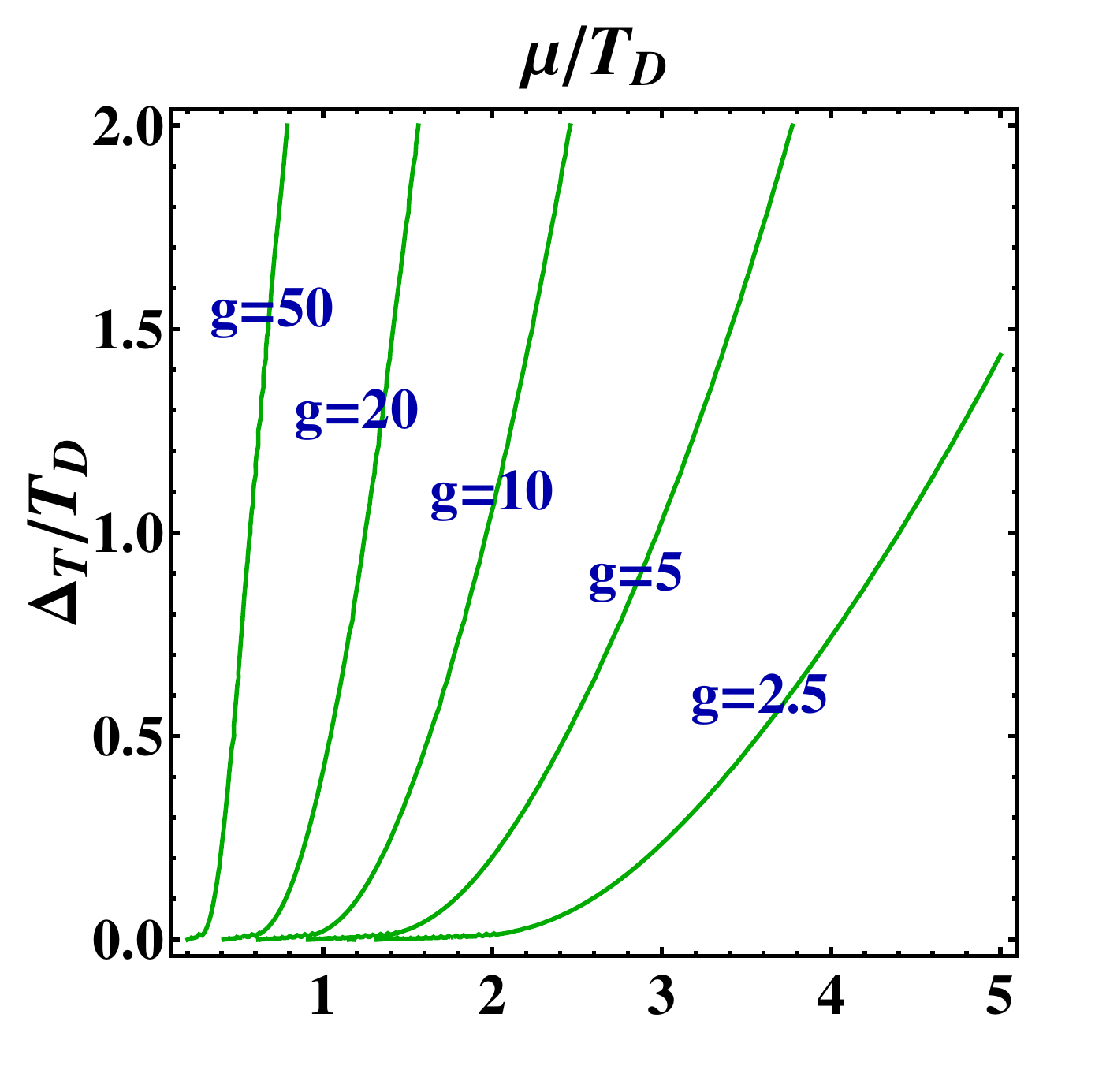}} \subfigure[]{%
\includegraphics[width=6cm]{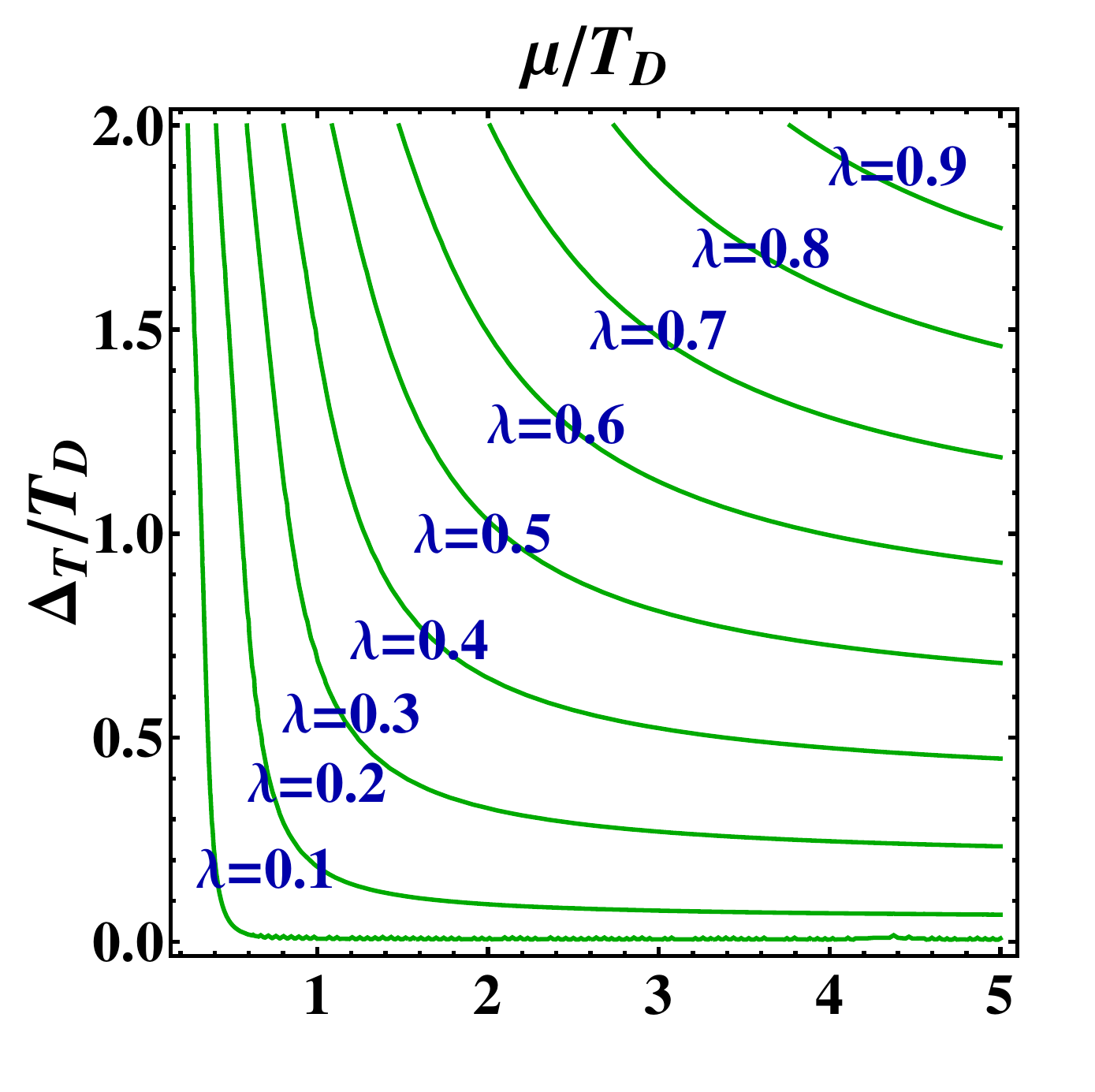}}
\caption{Triplet order parameter $\Delta _{T}$. (a) as function of $\protect%
\lambda $, (b) as function of $g$.}
\end{figure}

In several limiting cases the integrals can be performed analytically. At
zero chemical potential the results are presented in Section IV, while here
we list the BCS limit of $\mu >>T_{D}$ and the strong coupling case of $g\mu
^{2}>>1$,$\Delta _{T}\propto g$.

(i) In the BCS limit one has%
\begin{equation}
\frac{1}{g}=\frac{a_{T}\mu ^{2}}{4\pi ^{2}}\sinh ^{-1}\frac{T_{D}}{\Delta
_{T}}\text{,}  \label{gapBCS_T}
\end{equation}%
with $a_{T}=0.69$, leading to an exponential gap dependence on $\lambda $
when it is small:%
\begin{equation}
\Delta _{T}=T_{D}/\sinh \left( 1/2a_{T}\lambda \right) \simeq
2T_{D}e^{-1/2a_{T}\lambda }\text{.}  \label{dT_BCS}
\end{equation}

(ii) In the strong coupling one obtains with solution
\begin{equation}
\Delta _{T}=\frac{g}{12\pi ^{2}}\left\{
\begin{array}{c}
6\mu ^{2}+2\text{ \ for }\mu <1 \\
\left( \mu +1\right) ^{3}\text{\ for }\mu >1%
\end{array}%
\right. \text{,}  \label{dT_sc}
\end{equation}%
see Fig.4a. Usually the local coupling does not prefer the triplet pairing
and the singlet channels of coupling are realized. We therefore turn to them.

\subsection{Singlet representations.\textit{\ }}

It turns out that the second singlet in Eq.(\ref{singlets}) gives results
identical to that of the first one, while the third singlet does not have a
solution in the physically interesting range of parameters. Therefore we
assume the order parameter in the matrix form $\ \Delta =\Delta
_{S}M_{1}^{S}=i\Delta _{S}\alpha ^{y}$. The relevant ($41$) matrix element
of the matrix gap equation, Eq.(\ref{gap eq}), is for real $\Delta _{S}$:
\begin{figure}[tbp]
\centering
\subfigure[]{\includegraphics[width=6cm]{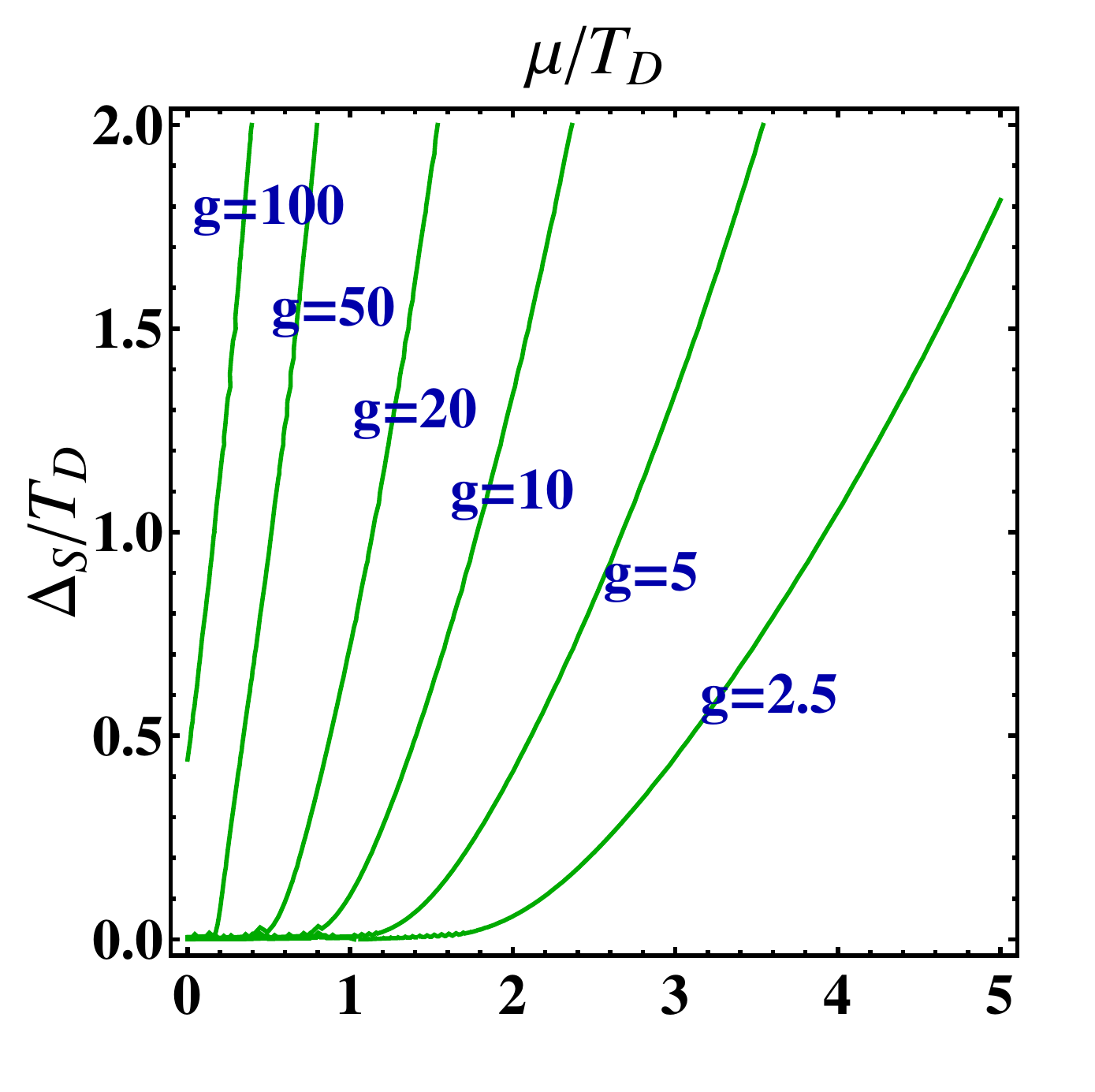}} \subfigure[]{%
\includegraphics[width=6cm]{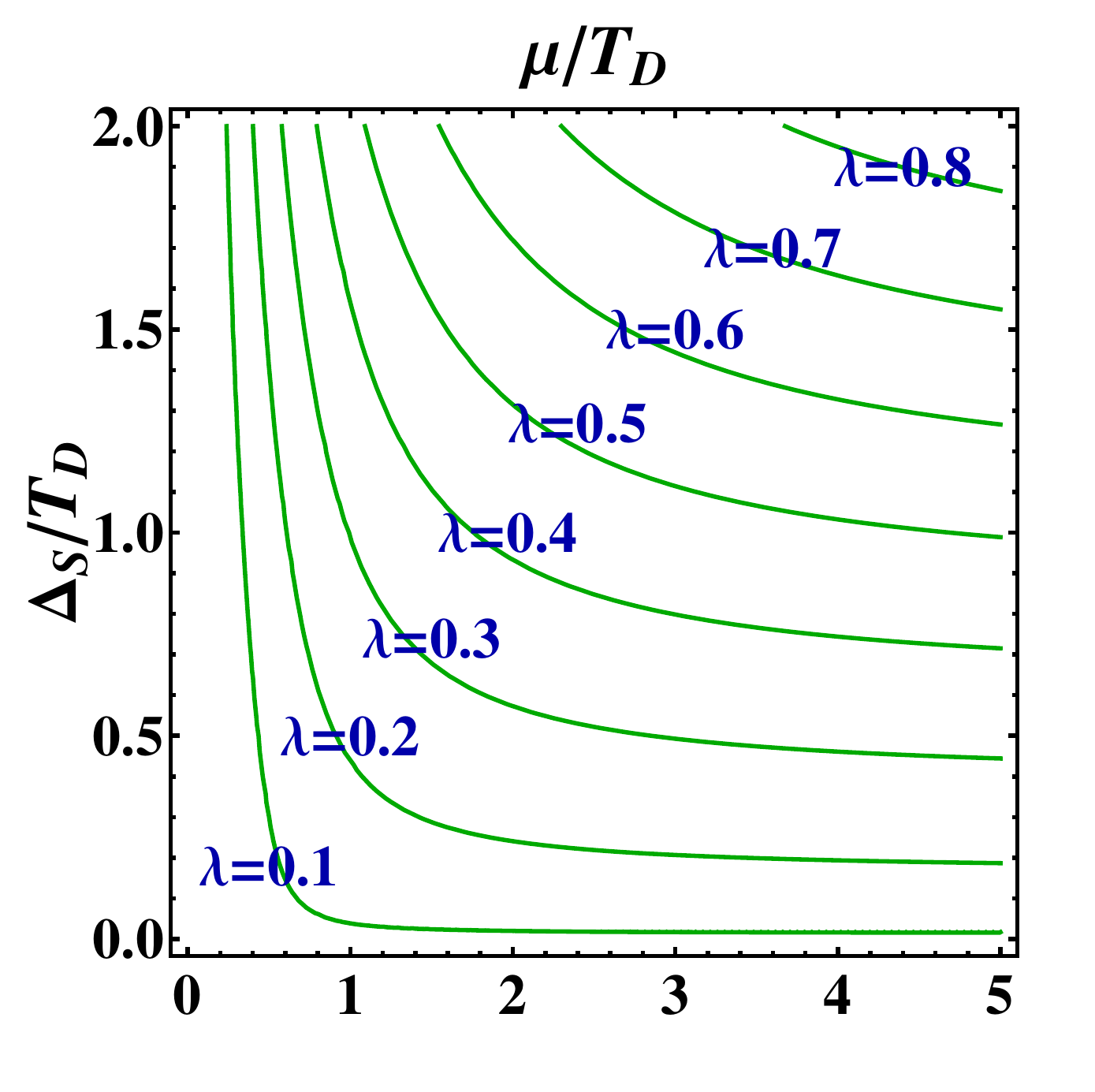}}
\caption{Singlet order parameter $\Delta _{S}$. (a) as function of $\protect%
\lambda $, (b) as function of $g$.}
\end{figure}

\begin{eqnarray}
\frac{1}{g} &=&\sum\limits_{\omega p}\left( \Delta _{S}^{2}+p^{2}+\mu
^{2}+\omega ^{2}\right) \times \left[ \left( \Delta _{S}^{2}+p^{2}\right)
^{2}\right.   \label{gapsinglet} \\
&&\left. +\left( \mu ^{2}+\omega ^{2}+2\Delta _{S}^{2}\right) \left( \mu
^{2}+\omega ^{2}\right) +2p^{2}\left( \omega ^{2}-\mu ^{2}\right) \right]
^{-1}\text{.}  \notag
\end{eqnarray}%
The spectrum (in physical units) now is isotropic,%
\begin{equation}
E_{\pm }^{2}=\Delta _{S}^{2}+\left( v_{F}\left\vert p\right\vert \pm \mu
\right) ^{2}\text{.}  \label{spectrum_S}
\end{equation}%
Integration over $\omega $ gives%
\begin{equation}
\frac{1}{g}=\mu \sum\limits_{\mu -T_{D}<\varepsilon _{p}<\mu +T_{D}}\frac{p}{%
r_{+}r_{-}\left( r_{+}-r_{-}\right) }\text{,}  \label{gap1_S}
\end{equation}%
where $r_{\pm }=\sqrt{\Delta _{S}^{2}+\left( \left\vert p\right\vert \pm \mu
\right) ^{2}}$, while the $p$ integration results in:

\begin{eqnarray}
\frac{16\pi ^{2}}{g} &=&\Phi \left( \mu +1\right) -\Phi \left( \max \left[
\mu -1,0\right] \right) \text{;}  \label{gap2_S} \\
\Phi \left( \mu \right) &=&r_{-}\left( p+3\mu \right) +r_{+}\left( p-3\mu
\right) -\left( \Delta _{S}^{2}-2\mu ^{2}\right)  \notag \\
&&\times \log \left[ \left( p+r_{-}-\mu \right) \left( p+r_{+}+\mu \right) %
\right] \text{.}  \notag
\end{eqnarray}%
The solution is presented in Fig. 5a and 5b as lines of constant $g$ and $%
\lambda ,$ respectively. One observes that the gaps are comparable to those
of the triplet shown in Fig.4 in whole range of parameters. The expression
for the gap simplifies for

(i) BCS, $\mu >>T_{D}$

\begin{equation}
\Delta _{S}=T_{D}/\sinh \left( 1/2\lambda \right) \simeq
2T_{D}e^{-1/2\lambda }\text{.}  \label{BCS_S}
\end{equation}

(ii) Strong coupling

\begin{equation}
\Delta _{S}=\frac{2\lambda \left( T_{D}+\mu \right) ^{3}}{3\mu ^{2}}\text{.}
\label{S_sc}
\end{equation}

Having found the order parameter, one has to determine what symmetry
breaking is realized by comparing energies of the solutions.

\section{Singlet vs triplet. Energy.}

We calculate the energy of a solution using the well known formula \cite{AGD}%
\begin{equation}
F=2\int_{\Delta ^{\prime }=0}^{\Delta }\frac{d\left( 1/g\right) }{d\Delta
^{\prime }}\Delta ^{\prime 2}\text{.}  \label{general_energy}
\end{equation}%
For the triplet and singlet solutions the result of integration performed
numerically is presented in Fig.6a. One observes that for all but the
smallest coupling $\lambda $ the channels are nearly degenerate although the
singlet is always lower. The lines of constant difference $F_{T}-F_{S}$ are
given in Fig. 6b as functions of $\mu $ and $\lambda $.
\begin{figure}[tbp]
\centering
\subfigure[]{\includegraphics[width=6cm]{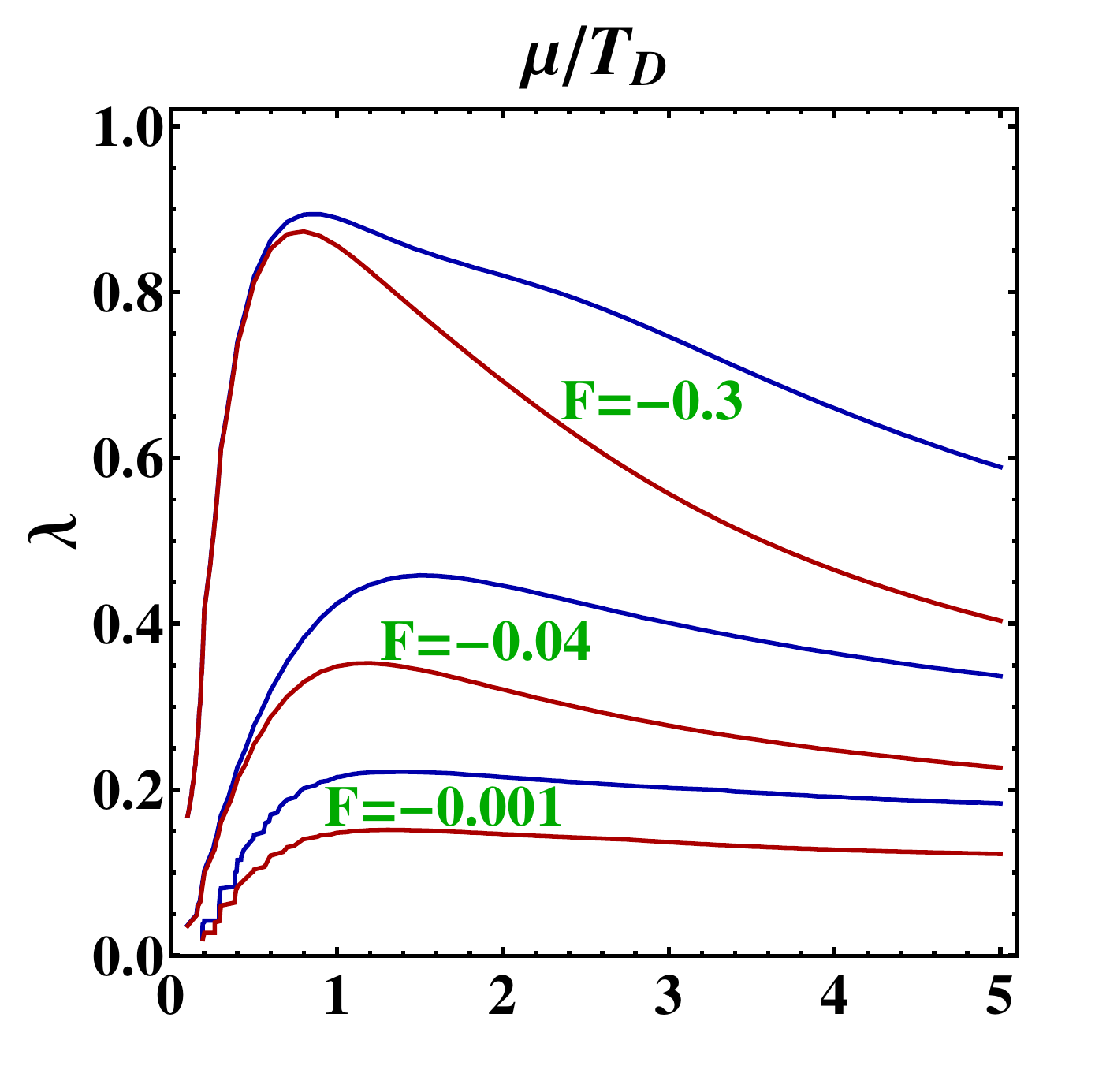}} \subfigure[]{%
\includegraphics[width=6cm]{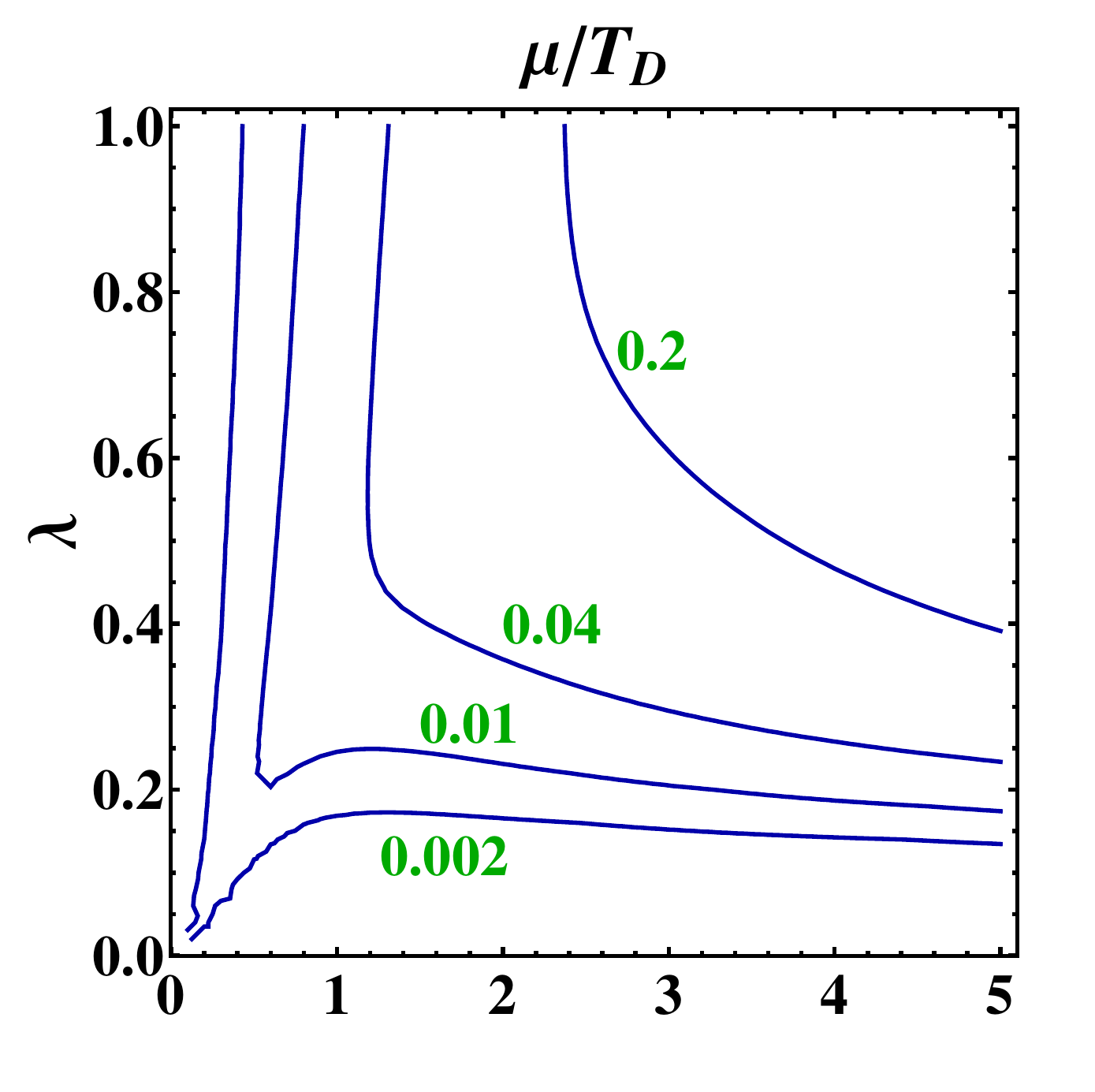}}
\caption{Energy of triplet and singlet. (a) Profile of constant energy for
singlet and triplet condensates in the $\protect\mu -\protect\lambda $
plane. (b) Difference $F_{T}-F_{S}$.}
\end{figure}
As can be seen, the difference becomes small especially at smaller chemical
potential. In Fig.7 more detailed results for the triplet and the singlet
order parameters are presented for chemical potential smaller than Debye
energy. One clearly observes the critical values of $8\pi ^{2}$ and $12\pi
^{2}$ for couplings $g$ when the singlet and triplet appears. They become
nearly degenerate just above $12\pi ^{2}$. Energies are also nearly
degenerate. The case of the quantum critical point $\mu =0$ is considered in
detail analytically in the following Section.
\begin{figure}[tbp]
\centering
\includegraphics[width=7cm]{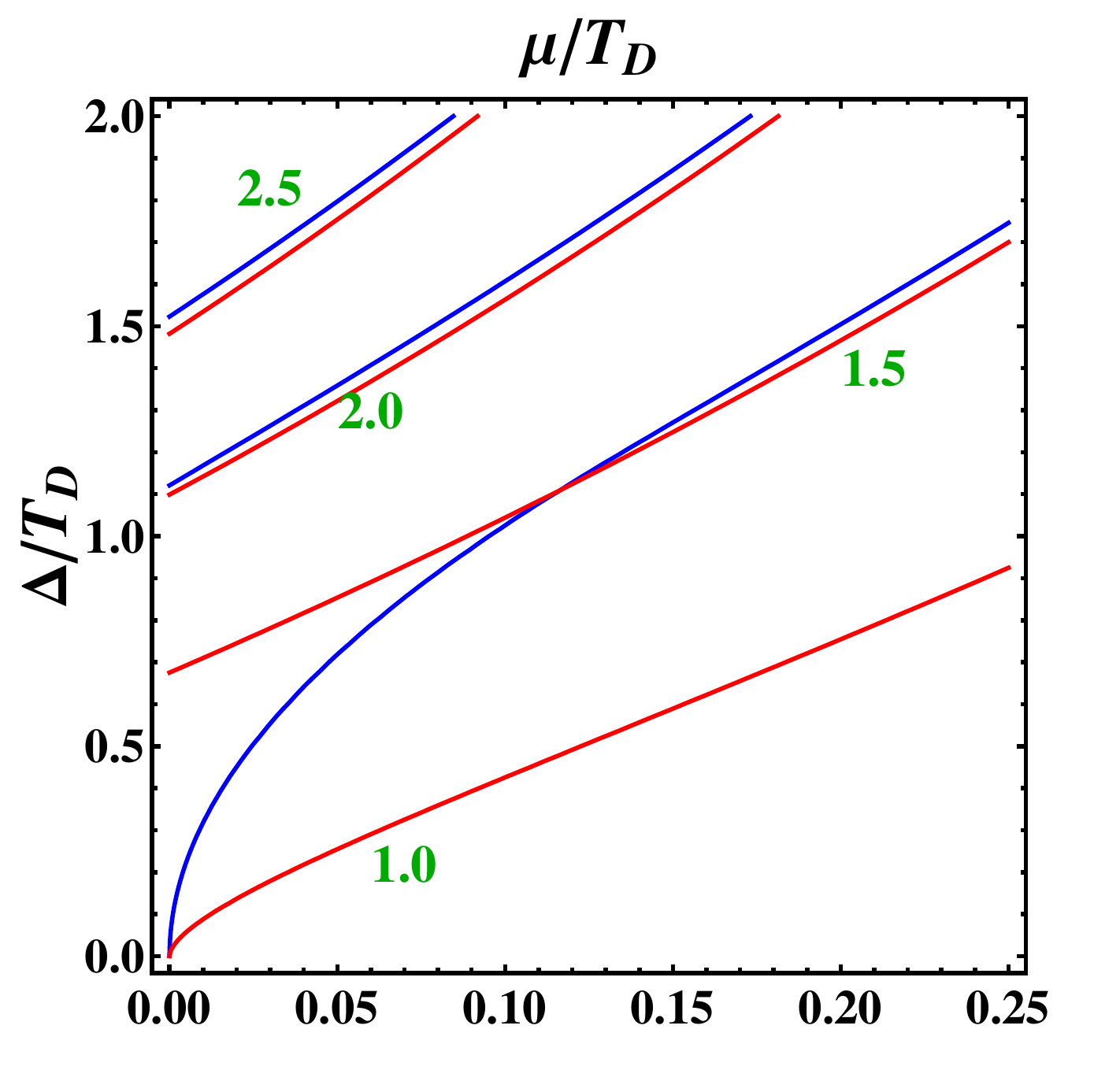} \vspace{-0.5cm}
\caption{Order parameters for triplet and singlet near criticality. Order
parameters of triplet (blue) and singlet at following values of coupling
constant (in units of critical value for singlet $g_{c}^{S}$): $1$, $3/2$
(critical value for triplet), $2$, $5/2$.}
\end{figure}

\begin{figure*}[tph]
\begin{center}
\subfigure[]{\includegraphics[width=5.8cm]{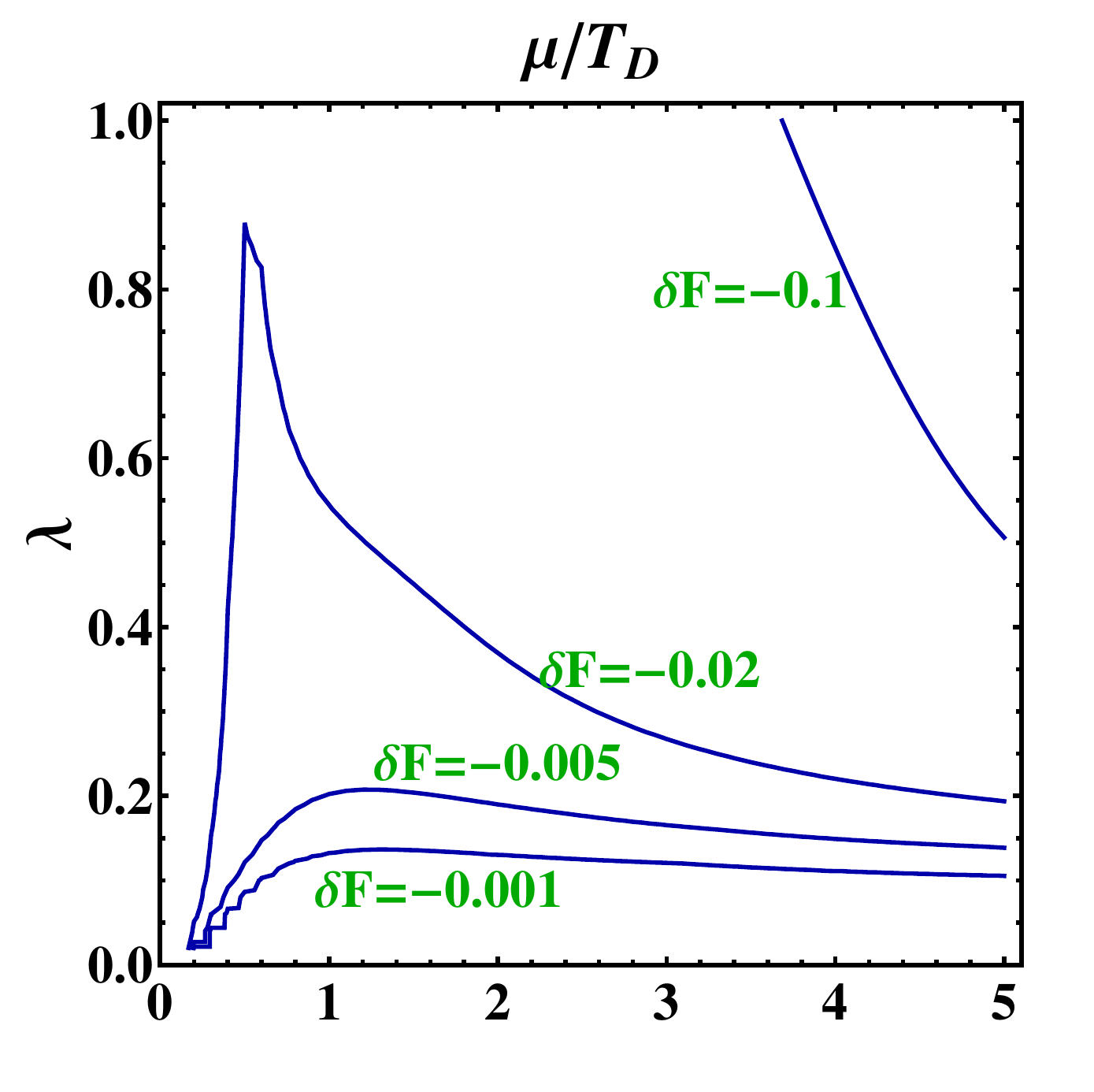}} \subfigure[]{%
\includegraphics[width=5.8cm]{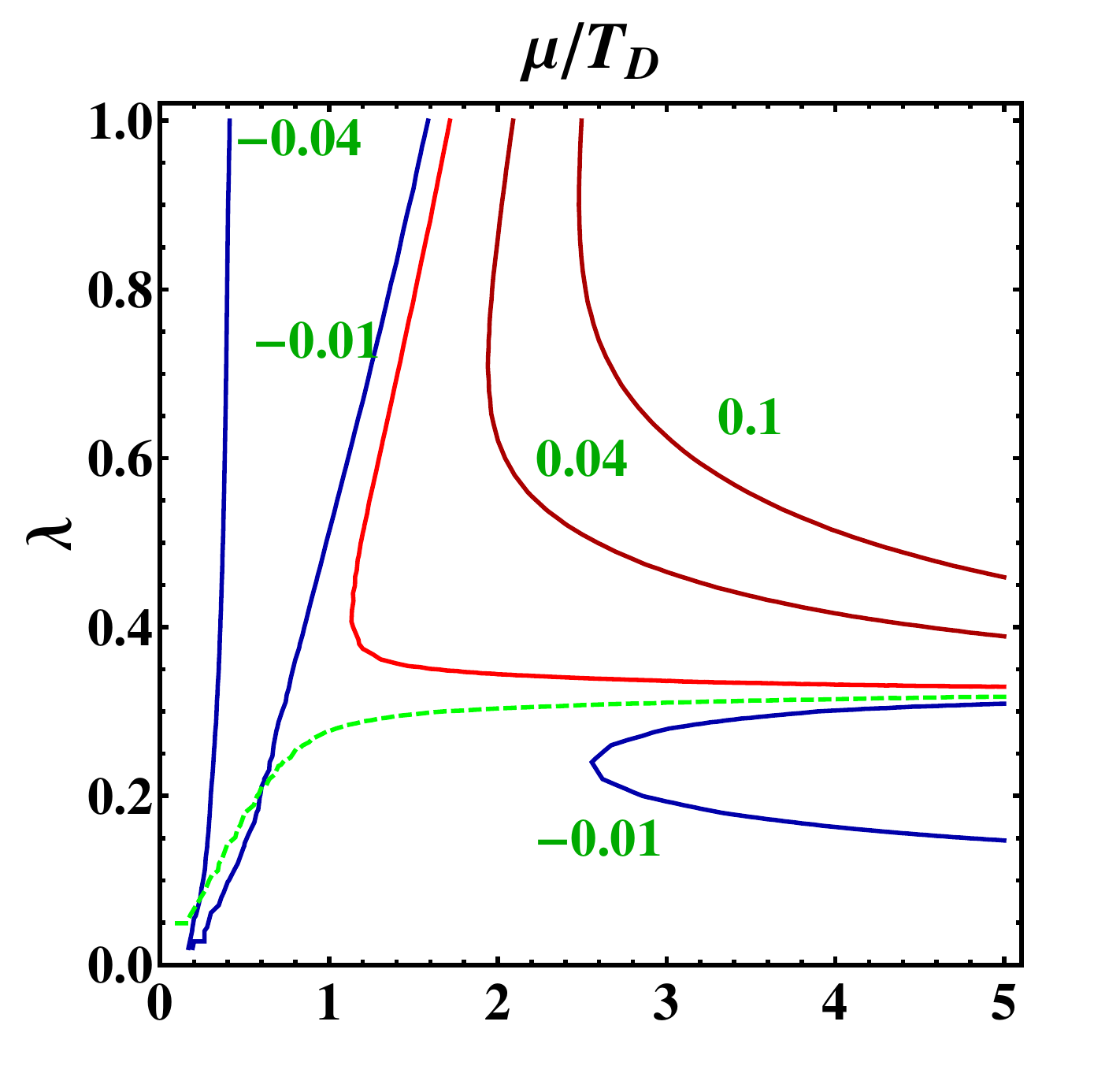}} \subfigure[]{%
\includegraphics[width=5.8cm]{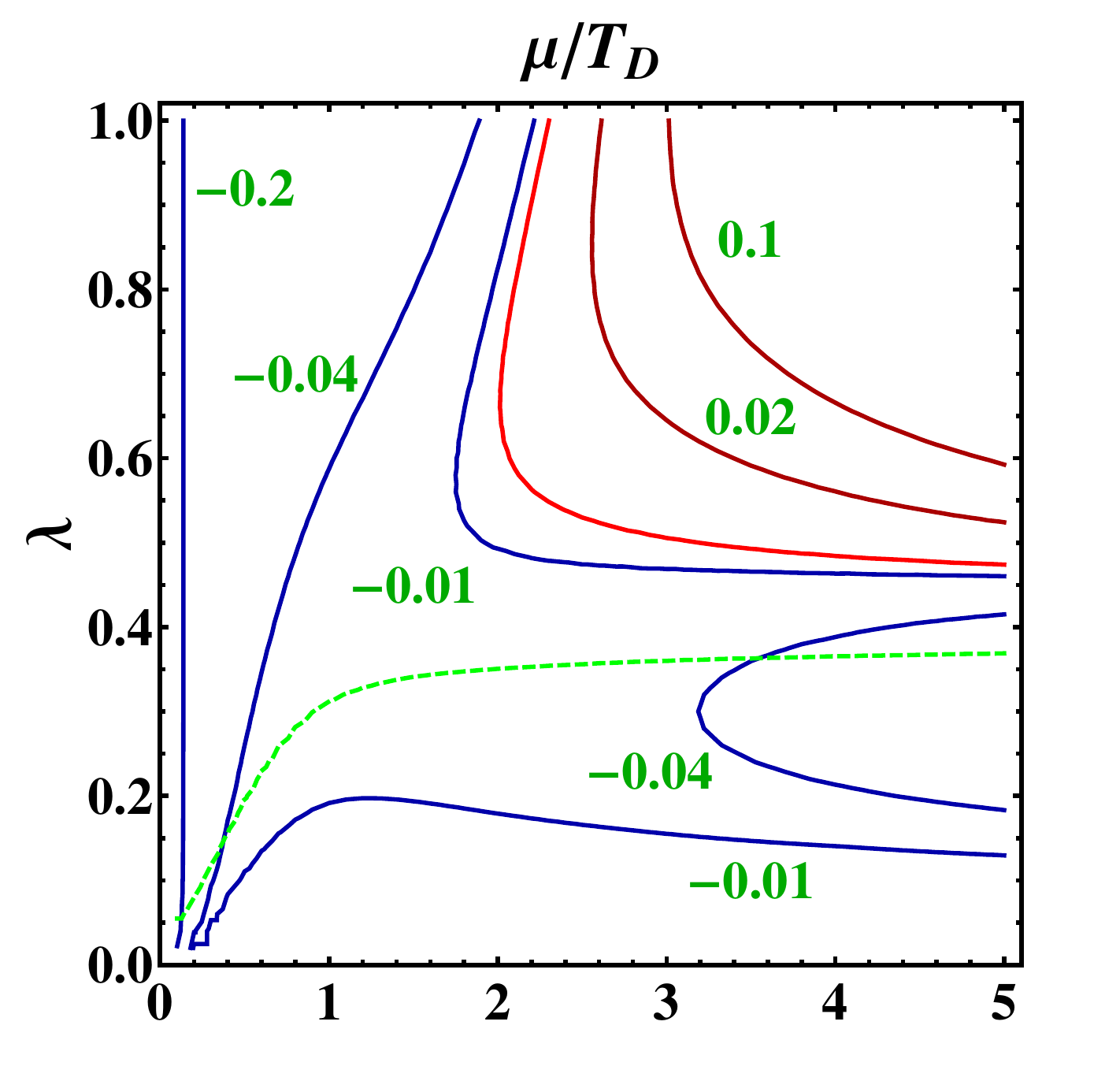}}
\end{center}
\par
\vspace{-0.5cm}
\caption{Comparison between energies of the singlet and triplet. (a)
Exchange corrections difference between the two channels: $\protect\delta F=$%
$\protect\delta F_{T}-\protect\delta F_{S}$. (b) Difference of energies
including the exchange correction for $\protect\lambda _{ex}=0.02$. The red
line separates the singlet phase from the triplet phase. Above the line on
the brown curves the energy difference $F_{T}-F_{S}$ is positive, while
below it on the green line it is negative. Portion of the phase diagram
below the dashed line requires consideration beyond perturbation theory. (c)
Same for much larger exchange coupling $\protect\lambda _{ex}=0.32$.}
\end{figure*}

In limiting cases, one obtains expressions in closed form.

(i) BCS, $\mu >T_{D}$, using Eq.(\ref{gapBCS_T}) and Eq.(\ref{dT_BCS}) for
the triplet and Eq.(\ref{BCS_S}) for the singlet, one has the energy density:%
\begin{eqnarray}
F_{T,S} &=&-\frac{a_{T,S}\mu ^{2}T_{D}}{2\pi ^{2}v_{F}^{3}\hbar ^{3}}\left(
\sqrt{\Delta _{T}^{2}+T_{D}^{2}}-T_{D}\right)  \label{F_BCS} \\
&\simeq &-\frac{a_{T,S}}{\pi ^{2}}\frac{\mu ^{2}T_{D}^{2}}{v_{F}^{3}\hbar
^{3}}\exp \left( -\frac{1}{a_{T,S}\lambda }\right) ,  \notag
\end{eqnarray}%
with $a_{T}=0.69$, while $a_{S}=1$ and assuming $\lambda <<1$. The ratio of
the two phases gives

\begin{equation}
\frac{F_{T}}{F_{S}}=0.69e^{-0.45/\lambda }\text{.}  \label{ratio}
\end{equation}

\bigskip (ii) Strong coupling limit, using Eq.(\ref{dT_sc}) for triplet and
Eq.(\ref{S_sc}) for the singlet,

\begin{equation}
F_{T}=F_{S}=-\frac{1}{72\pi ^{4}v_{F}^{3}\hbar ^{3}}\left\{
\begin{array}{c}
4\left( 3\mu ^{2}+T_{D}^{2}\right) ^{2}\text{ \ for }\mu <T_{D} \\
T_{D}^{-2}\left( \mu +T_{D}\right) ^{6}\text{\ \ for }\mu >T_{D}%
\end{array}%
\right. \text{.}  \label{energy_sc}
\end{equation}%
The difference appears at order $1/g$.

To summarize, in most of the parameter range shown triplet is a bit higher
than that of the singlet, but the two condensates are nearly degenerate. The
degeneracy in practise is lifted in favor of the triplet by the spin-spin
interaction, Eq.(\ref{s-s}), or magnetic impurities present in materials
exhibiting the 3D Dirac point.

\subsection{The influence of exchange.}

Let us estimate the perturbatively the energy change due to the exchange
interactions due to Stoner exchange. In the simplest case of local spin
attraction one uses the Stoner approximation\cite{White}, $J\left( \mathbf{r}%
\right) =I\delta \left( \mathbf{r}\right) $, where $I$ is the Stoner
constant, using the Gaussian factorization one obtains

\begin{eqnarray}
\delta F &=&-\frac{I}{2V}\int_{\mathbf{r}}\Sigma _{\alpha \beta }^{i}\Sigma
_{\gamma \delta }^{i}\left\langle \psi _{\alpha }^{+}\left( r\right) \psi
_{\beta }\left( r\right) \psi _{\gamma }^{+}\left( r\right) \psi _{\delta
}\left( r\right) \right\rangle  \notag \\
&\simeq &\frac{I}{2V}\int_{\mathbf{r}}\Sigma _{\alpha \beta }^{i}\Sigma
_{\gamma \delta }^{i}\left\langle \psi _{\alpha }^{+}\left( r\right) \psi
_{\gamma }^{+}\left( r\right) \right\rangle \left\langle \psi _{\beta
}\left( r\right) \psi _{\delta }\left( r\right) \right\rangle  \notag \\
&=&\frac{I}{2g^{2}}\Delta _{\gamma \alpha }^{\ast }\Sigma _{\alpha \beta
}^{i}\Delta _{\beta \delta }\Sigma _{\delta \gamma }^{it}\text{.}
\label{deltaF}
\end{eqnarray}%
The triplet, $\Delta =\Delta _{T}\beta \alpha ^{x}$, predictably gains energy

\begin{equation}
\delta F_{T}=-\frac{2I}{g^{2}}\Delta _{T}^{2}\text{,}  \label{deltaT}
\end{equation}%
while singlet, $\Delta =i\Delta _{S}\alpha ^{y}$ loses energy

\begin{equation}
\delta F_{S}=\frac{6I}{g^{2}}\Delta _{S}^{2}\text{.}  \label{deltaS_}
\end{equation}

As in the case of the phonon induced interactions, a more convenient
dimensionless quantity describing the exchange is
\begin{equation}
\lambda _{ex}=ID\left( \mu \right) =I\mu ^{2}/\left( 8\pi ^{2}v_{F}^{3}\hbar
^{3}\right) \text{.}  \label{lambdaex}
\end{equation}%
We assume that the value is quite far from the Stoner ferromagnetic
instability value ($\lambda _{ex}=1$). The gain of triplet over the singlet
is therefore written as%
\begin{equation*}
\delta F_{T}-\delta F_{S}=-\frac{2\lambda _{ex}}{\lambda ^{2}}D\left( \mu
\right) \left( \Delta _{T}^{2}+3\Delta _{S}^{2}\right) \text{,}
\end{equation*}%
and is given in Fig.8a. The difference of full energies is given in
Figs.8b-8c for two values of the exchange coupling. The crossover from
singlet to triplet occurs, $F_{T}+\delta F_{T}=F_{S}+\delta F_{S}$ at the
following value of the exchange coupling:%
\begin{equation}
\lambda _{ex}^{c}=\frac{\lambda ^{2}}{2D\left( \mu \right) }\frac{F_{T}-F_{S}%
}{3\Delta _{S}^{2}+\Delta _{T}^{2}}\text{.}  \label{Jcross}
\end{equation}%
In the general case two values of exchange coupling were calculated
numerically leading to the phase diagram shown in Fig. 9. General feature of
the phase diagram is that the triplet superconductivity might appears either
at small chemical potential or at very large one compared to $T_{D}$. The
second possibility is not realized. Since perturbation theory in exchange
coupling was used, the estimate is valid only when $F_{S,T}>>\delta F_{S,T}$
marked by dashed lines on Fig.8b,c. On the lines the perturbation is half of
the leading order. We argue that in this region either a ferromagnetic state
is formed or the perturbation theory is not valid. In limiting cases
analytic expression can be obtained.
\begin{figure}[t]
\centering
\includegraphics[width=7cm]{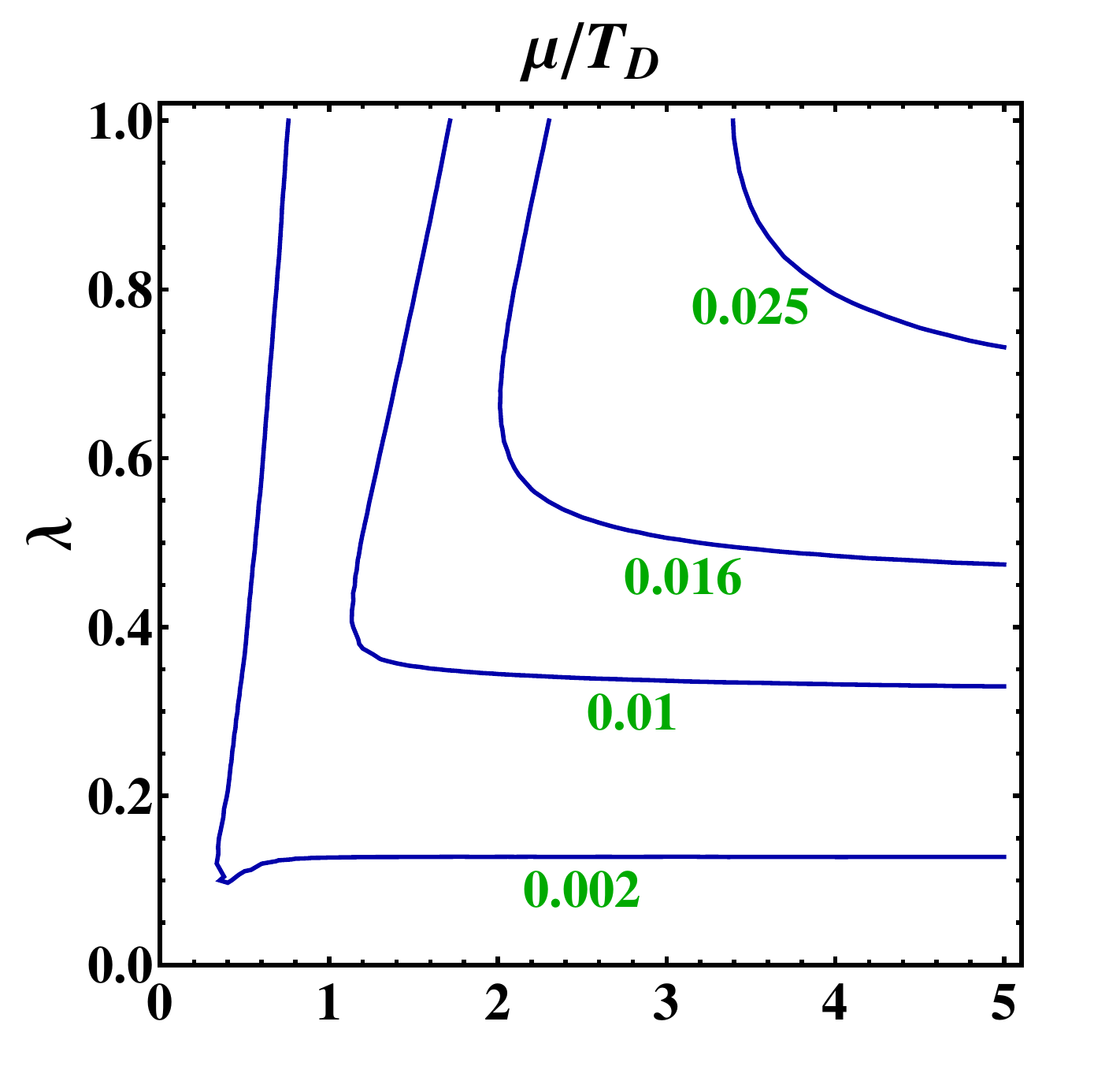} \vspace{-0.5cm}
\caption{Critical exchange coupling for various chemical potential $\protect%
\mu $ and the phonon induced electron-electron coupling $\protect\lambda $.}
\end{figure}

(i) For $\mu >>T_{D}$ according to Eq.(\ref{F_BCS})

\begin{eqnarray}
F_{T}+\delta F_{T} &=&-8D\left( \mu \right) T_{D}^{2}\left( a_{T}+\frac{%
\lambda _{ex}}{\lambda ^{2}}\right) e^{-1/a_{T}\lambda };  \label{enBCS} \\
F_{S}+\delta F_{S} &=&-8D\left( \mu \right) T_{D}^{2}\left( 1-3\frac{\lambda
_{ex}}{\lambda ^{2}}\right) e^{-1/\lambda }\text{.}  \notag
\end{eqnarray}%
Therefore the transition occurs when
\begin{equation}
\lambda _{ex}^{c}=\lambda ^{2}\frac{e^{\left( a_{T}^{-1}-1\right) /\lambda
}-a_{T}}{3e^{\left( a_{T}^{-1}-1\right) /\lambda }+1}\approx \frac{\lambda
^{2}}{3}\text{.}  \label{g0BCS}
\end{equation}

(ii) In the strong coupling for $\mu <<T_{D}$ , $\Delta _{T}=\Delta _{S}\sim
\frac{T_{D}^{4}}{18\pi ^{4}v_{F}^{3}\hbar ^{3}}$, so that the difference is
\begin{equation}
\delta F_{T}-\delta F_{S}\sim -\frac{32}{9\pi ^{2}}\frac{\lambda
_{ex}T_{D}^{4}}{\mu ^{2}}\text{.}  \label{deltaEstrong}
\end{equation}%
The triplet is always favored in this limit due to degeneracy of energies
without the exchange coupling.

\bigskip As is seen from figure 8 the most promising region in parameter
space in which the triplet superconductivity prevails is at small chemical
potential. As was mentioned in Introduction, the "extreme" case of zero
chemical potential can be physically achieved by tuning parameters of the
material to the transition between the topological insulator phase and the
band insulator phase, so we study it in more detail.

\section{Quantum critical point at zero chemical potential and its critical
exponents.}

A peculiarity of superconductivity in Dirac semimetal at zero chemical
potential is that electrons (and holes) in Cooper pairs are created
themselves by the pairing interaction rather than being present in the
sample as free electrons. Therefore it is shown that it is possible to
neglect the effect of weak doping and consider directly the $\mu =0$
particle-hole symmetric case. This point in parameter space is the QCP \cite%
{Sachdev}. Microscopically, Cooper pairs of both electrons and holes are
formed. The system is unique in this sense since the electron - hole
symmetry is not spontaneously broken in both normal and superconducting
phases. Supercurrent in such a system does not carry momentum or mass. We
discuss the triplet state followed by the singlet.
\begin{figure}[tbp]
\centering
\subfigure[]{\includegraphics[width=6cm]{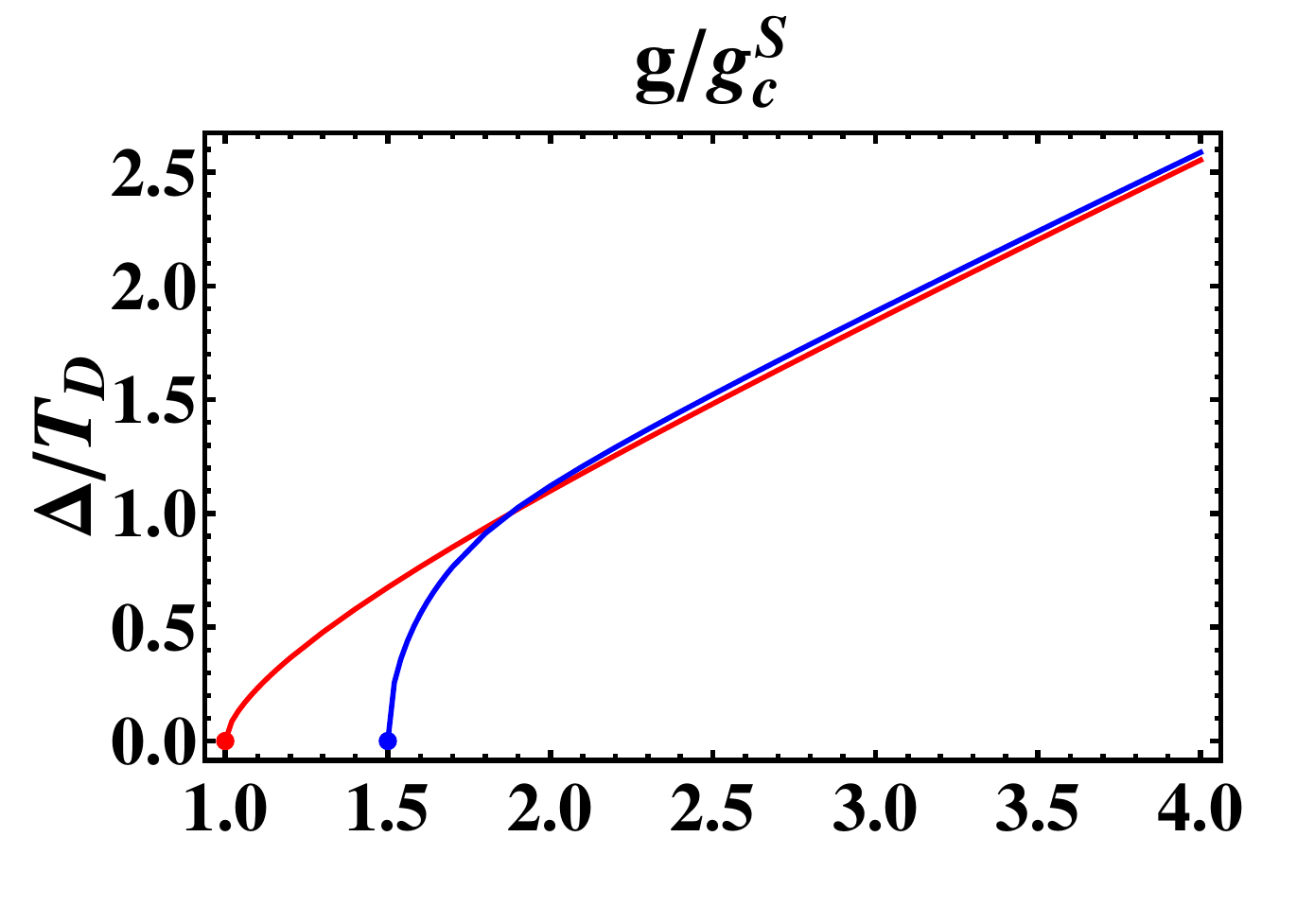}} \subfigure[]{%
\includegraphics[width=6cm]{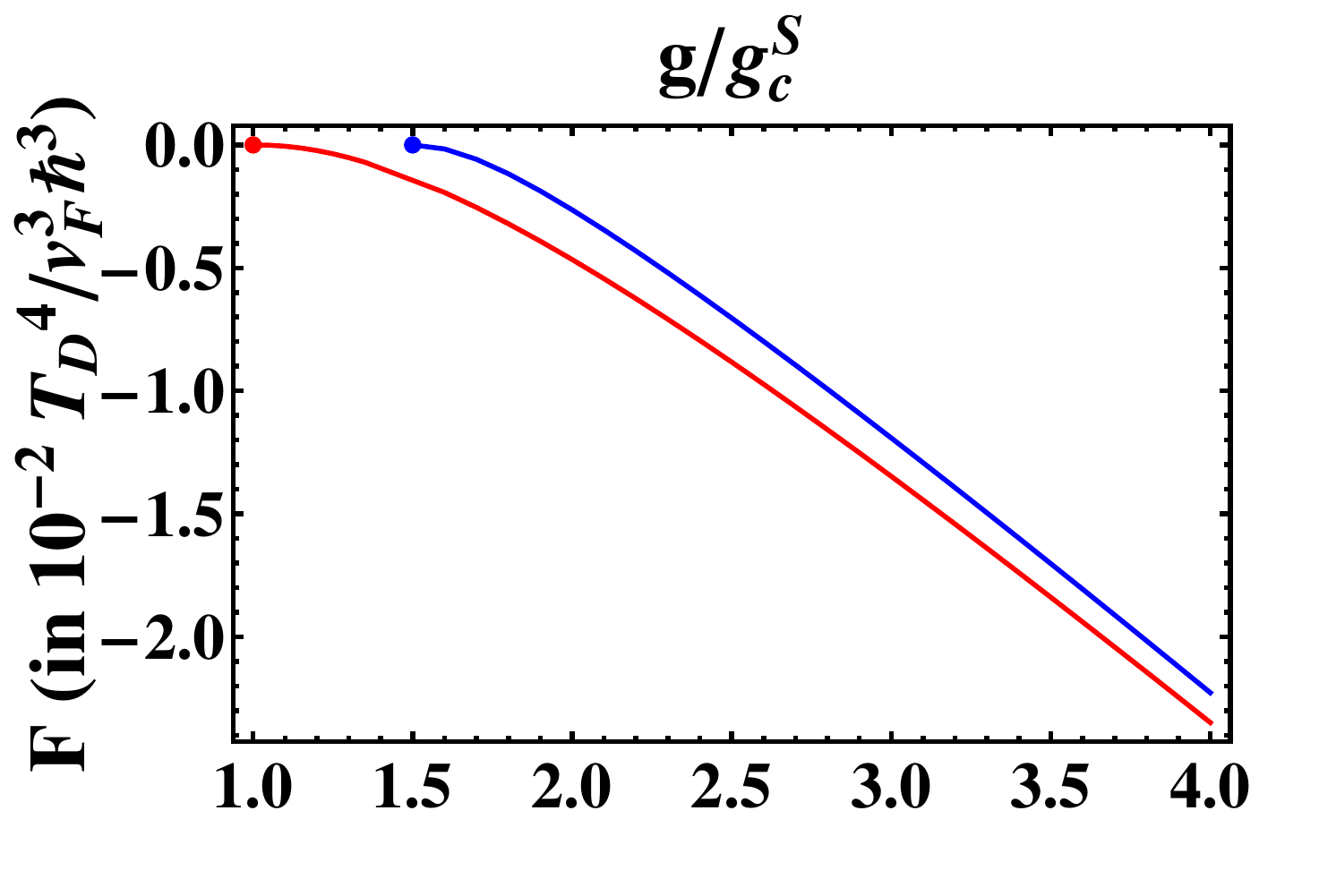}}
\caption{Quantum critical point $\protect\mu =0$. (a) Singlet and triplet
order parameter as function of $g$. (b) Energies.}
\end{figure}

\subsection{Triplet.}

Spectrum of the triplet becomes very simple, $E_{\pm }^{2}=\left( \Delta
_{T}\pm v_{F}\left\vert p_{z}\right\vert \right) ^{2}+v_{F}^{2}p_{\perp
}^{2} $. Performing analytically the integral over the angle and momentum in
the gap equation, Eq.(\ref{gapeqT}), one obtains

\begin{equation}
\frac{12\pi ^{2}v_{F}^{3}\hbar ^{3}}{g}=\left\{
\begin{array}{c}
T_{D}^{2}-\frac{\Delta _{T}^{2}}{5}\text{ for }\Delta _{T}<T_{D} \\
\frac{T_{D}^{3}}{\Delta _{T}}-\frac{T_{D}^{5}}{5\Delta _{T}^{3}}\text{ for }%
\Delta _{T}>T_{D}%
\end{array}%
\right. \text{.}  \label{gapQCP}
\end{equation}%
The solution of the equation for $\Delta _{T}$ as function of coupling $g$
is presented in Fig. 10a. The triplet superconducting solution exists, like
in the 2D case\cite{Li14}, only when the coupling exceeds a critical value
(in physical units),%
\begin{equation}
g_{c}^{T}=12\pi ^{2}\frac{v_{F}^{3}\hbar ^{3}}{T_{D}^{2}}\text{.}  \label{gc}
\end{equation}%
The dependence on the cutoff $T_{D}$ is incorporated in the renormalized
coupling with dimension of energy defined as
\begin{equation}
U_{T}^{2}=T_{D}^{2}\left( 1-\frac{g_{c}^{T}}{g}\right) \text{.}  \label{Udef}
\end{equation}%
This quantity can be interpreted as an effective binding energy of the
Cooper pair in the Dirac semi - metal. The dependence of the gap is $\Delta
_{T}=\sqrt{5}U_{T}$ for $U_{T}<5^{-1/2}$ (or $g<5/4g_{c}^{T}$). The critical
exponent therefore is $\Delta _{T}\propto \left( 1-g_{c}^{T}/g\right)
^{\beta }$ for $\beta =1/2$. This defines the (zero temperature) triplet
quantum critical point.

Energy, calculated using the AGD formula, Eq.(\ref{general_energy}), can be
written via the energy gap in a closed form:%
\begin{equation}
F_{T}=-\frac{T_{D}^{2}}{5g_{c}^{T}}\left\{
\begin{array}{c}
\frac{\Delta _{T}^{4}}{T_{D}^{4}}\text{ for }\Delta _{T}<T_{D} \\
10\frac{\Delta _{T}}{T_{D}}-15+\frac{6T_{D}}{\Delta _{T}}\text{ for }\Delta
_{T}>T_{D}%
\end{array}%
\right. \text{.}  \label{FTsmall}
\end{equation}%
Near criticality, Eq.(\ref{FTsmall}), $F_{T}\propto \left(
1-g_{c}^{T}/g\right) ^{2-\alpha }$, determines the quantum critical exponent
$\alpha =2$. Critical exponents coincide with the classical mean field 3D
exponents.

In the strong coupling limit $\Delta _{T}=T_{D}g/g_{c}^{T}$ and $%
F_{T}=-2gT_{D}^{2}/\left( g_{c}^{T}\right) ^{2}$. As we see in the next
subsection, the triplet QCP is unstable since the singlet order parameter
solution has lower energy.

\subsection{Singlet.\textit{\ }}

The spectrum is relativistic with the rest mass equal to the gap, $%
E^{2}=\Delta _{S}^{2}+v_{F}^{2}p^{2}$. The gap equation after integrations is%
\begin{equation}
\frac{8\pi ^{2}v_{F}^{3}\hbar ^{3}}{g}=T_{D}\sqrt{\Delta _{S}^{2}+T_{D}^{2}}%
-\Delta _{S}^{2}\sinh ^{-1}\left( T_{D}/\Delta _{S}\right) \text{.}
\label{gapeq_QCP_S}
\end{equation}%
The critical value is therefore lower than that for the triplet%
\begin{equation}
g_{c}^{S}=8\pi ^{2}\frac{v_{F}^{3}\hbar ^{3}}{T_{D}^{2}}\text{.}  \label{gcS}
\end{equation}%
In terms of the renormalized coupling, $U_{S}^{2}=T_{D}^{2}\left( 1-\frac{%
g_{c}^{S}}{g}\right) $, the gap equation near criticality takes the form%
\begin{equation}
U_{S}^{2}=\Delta _{S}^{2}\log \left( \frac{2T_{D}}{\sqrt{e}\Delta _{S}}%
\right) \text{.}  \label{deviationS}
\end{equation}%
The solution of Eq.(\ref{gapeq_QCP_S}) is given in Fig. 10a (red curve).

At small deviations from criticality one can approximate solution as $\Delta
_{S}=U_{S}\log ^{-1/2}\left( 2T_{D}/\sqrt{e}U_{S}\right) $, while in the
strong coupling limit, $\Delta _{S}=\frac{2g}{3g_{c}^{S}}T_{D}$. The energy
is

\begin{equation}
F_{S}=\frac{2T_{D}}{g_{c}^{S}}\left( T_{D}-\sqrt{\Delta _{S}^{2}+T_{D}^{2}}%
\right) +\frac{1}{g}\Delta _{S}^{2}\text{.}  \label{FS_QCP}
\end{equation}%
Near critical coupling, $F_{S}\simeq -\frac{U_{S}^{4}}{T_{D}^{2}g_{c}^{S}}%
\log ^{-1}\left( 2T_{D}/\sqrt{e}U_{S}\right) $, while in the strong coupling
limit one obtains again degeneracy with the triplet, $%
F_{S}=F_{T}=-8T_{D}^{2}g/\left( 3g_{c}^{S}\right) ^{2}$, see Fig.10b,
consistent with the general chemical potential result.

\subsection{The singlet triplet crossover due to exchange interaction.}

When the exchange interaction is added perturbatively (at coupling above the
critical one for the triplet), the energies of the competing condensates are
shifted; the crossover exchange (Stoner) coupling constant $I,$ defined in
Eq.(\ref{delta}), is given in Fig. 10 as function of the electron - electron
local coupling $g$. For $g$ just above the critical for triplet $g_{c}^{T}$,
see Eq.(\ref{gc}), the value of the $I_{c}$ is about $I_{c}=6$ (in units of $%
v_{F}^{3}\hbar ^{3}/T_{D}^{2}$), so that $I_{c}=\frac{6}{12\pi ^{2}}%
=\allowbreak 0.\,\allowbreak 05$. As the phonon mediated attraction strength
increases the critical value of exchange decreases as $1/g$.
\begin{figure}[tbp]
\centering
\includegraphics[width=7cm]{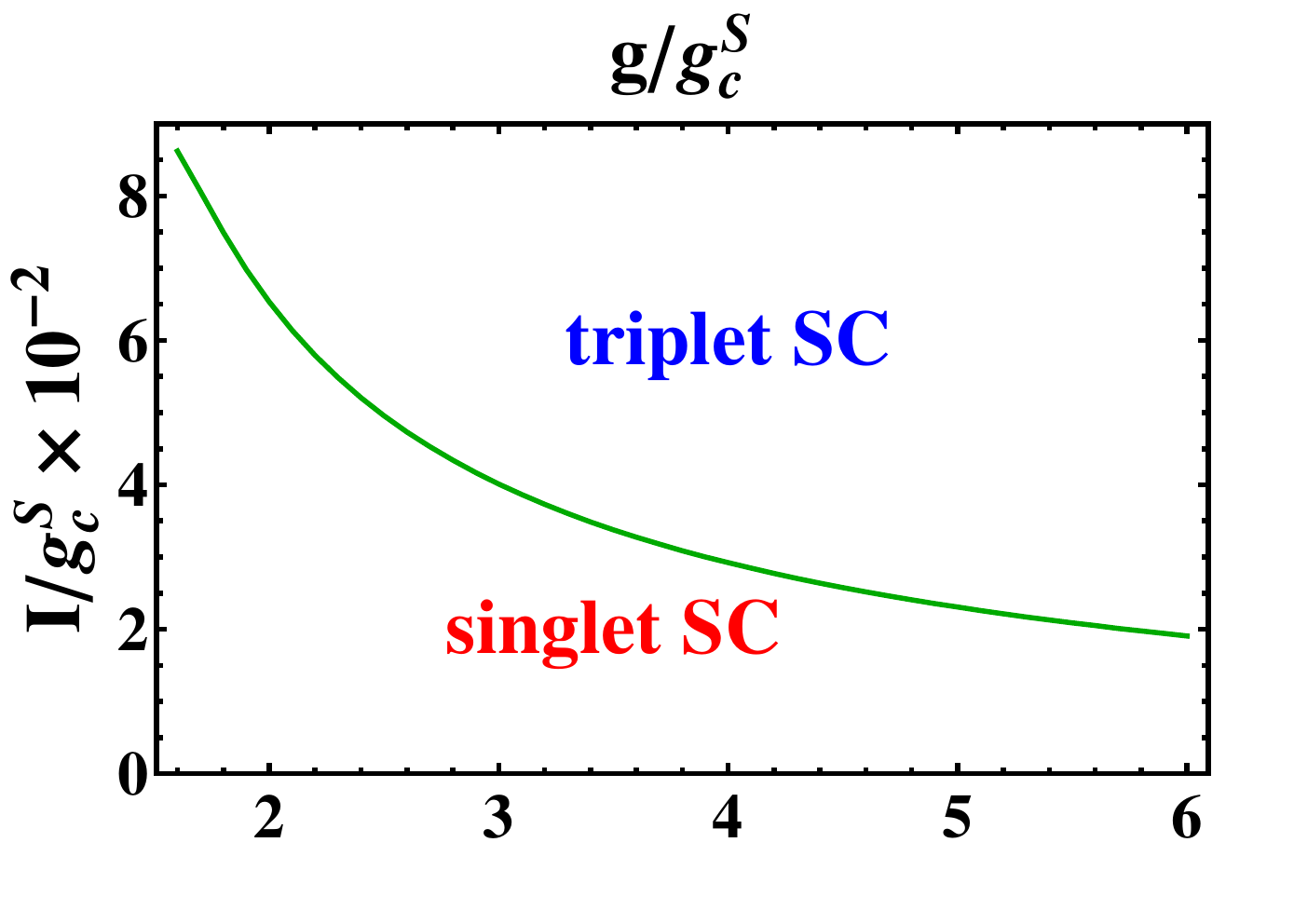} \vspace{-0.5cm}
\caption{Critical exchange coupling as function of $g$ at quantum critical
point.}
\end{figure}
The Dirac superconductor therefore is a rare example of 3D quantum critical
point.

\section{Discussion and conclusions.Summary.\textit{\ }}

To summarize, we have constructed a microscopic theory of superconductivity
(at zero temperature) in 3D time reversal and inversion invariant massless
Dirac{\LARGE \ }semi - metals. In these materials there are at least two
bands of opposite chiralities. Such a band structure appear in many recently
studied materials including copper doped TI $Bi_{2}Se_{3}$ in which the
triplet superconductivity is suspected \cite{pressureBiSe}.

In the framework of the "conventional" phonon mediated local attraction
model we classified, under simplifying assumptions of the 3D rotation
invariance, inversion and the time reversal, possible pairing channels.
There are three singlet channels and one triplet. Comparison of energies of
these condensates for arbitrary chemical potential and the electron-electron
interaction strength demonstrates that a singlet pairing always prevails, as
shown in Fig. 5. However, one notices that in large portions of the phase
diagram the energy density differences are much smaller that the typical
values of energy densities themselves. This means that interactions that are
small but discriminate between the spin singlet and the spin triplet are
important in order to determine the nature of the superconducting order
there.

The best candidate for such an interaction in materials under consideration
is the exchange (the Stoner term). Parameters of the model are therefore the
chemical potential $\mu $, the effective electron - electron coupling
strength $g$ and the Stoner coupling exchange constant $I$. Our main results
are given in Figs. 8a, 8b and 11. In certain ranges of parameters that
include the electron - phonon coupling parametrized by dimensional effective
electron - electron coupling $\lambda $ and the exchange interaction
parametrized by $\lambda _{ex}$, the triplet pairing is favored over the
singlet one. Fig. 8a, 8b demonstrate that the triplet exists either at small
chemical potential of order Debye energy $T_{D}$ or perhaps as small $%
\lambda $ and large chemical potential, while the singlet prevails in the
upper-right corner of the diagram beyond the red line.

The second region where triplet is competitive happens to be beyond the
range of validity of the perturbation theory and in fact will not
materialize since the superconducting order instability is probably weaker
than the Stoner instability for ferromagnetic correlations, so we are left
with the triplet states when the chemical potential is small.

To this end we have investigated the limit of zero chemical potential, where
the tendency towards the triplet pairing should be maximal. This is
presented in Fig.11. In this limit one cannot use the dimensionless coupling
strengths $\lambda $ and $\lambda _{ex}$, therefore should go back to the
dimensional coupling strengths $g$ and $I\,\ $related to the former by Eqs.(%
\ref{lambda},\ref{lambdaex}) used in this phase diagram. Transition to
superconductivity in this case is a rare occurrence of quantum critical
point in 3D with distinct critical couplings and exponents.

\subsection{Experimental feasibility of the triplet superconductivity due to
phonon and exchange interactions.\textit{\ }}

To estimate the pairing efficiency due to phonons, one should rely on recent
studies \cite{DasSarma14}. The effective dimensionless electron - electron
coupling constant due to phonons $\lambda ,$ defined in Eq.(\ref{lambda}),
is obtained from the exchange of acoustic phonons and is of order\cite{Cava}
$0.1-1$ (somewhat lower values are obtained in ref.\cite{Guinea}). Note that
a reasonable electron density of $n=3\cdot 10^{11}cm^{-2}$ in $Bi_{2}Te_{3}$
already conforms to the requirement that chemical potential less than the
Debye cutoff energy $T_{D}=300K$.

To estimate the strength of the exchange interactions due to itinerant
electrons one, as usual, starts from the Coulomb repulsion. The effects of
Coulomb interaction in 3D Dirac electrons are being studied extensively\cite%
{Wan}. RG analysis reveals the logarithmic divergence of Fermi velocity $%
v_{F}$, while the effective fine structure constant $\alpha =e^{2}/\hbar
v_{F}$ is marginally irrelevant. When a Dirac point is located on the Fermi
level, the Coulomb interaction is not screened. The Stoner theory\cite{White}
predicts that when $\lambda _{ex}$ becomes of order $1$, the material
develops ferromagnetism. Below that only the correlations play a role, but
as is seen in Figs.8a-b, such a relatively small exchange is sufficient to
damage the singlet condensate in favor of the triplet.

\subsection{Feasibility of observing the quantum criticality.}

In this paper we especially focused on the qualitatively distinct case of
Dirac fermions with small chemical potential. The situation is quite similar
to that of the 2D Weyl semi-metal in topological insulators. Although in the
original proposal of TI in materials \cite{Zhang1} the chemical potential
was zero, in experiments one finds often that the Dirac point is shifted
away from the Fermi surface by a significant fraction of $eV$ \cite{Zhang}.
There are however experimental methods to shift the location of the point by
doping (for example by copper\cite{Ong}), gating, pressure etc.\cite{bias}.
Superconductivity was in fact observed in otherwise non-superconducting TIs $%
Bi_{2}Te_{3}$ and $Bi_{2}Se_{3}$ under pressure\cite{pressureBiSe} $%
v_{F}\approx 7\ \cdot 10^{5}m/s$ (for $Bi_{2}Se_{3}$). It is possible that
at a certain pressure the system passes through the quantum critical point
and is therefore a candidate for the maximal enhancement of the triplet
superconductivity.

\subsection{Possible generalizations and comparison with other works.}

Here we compare our results with the earlier work ref.\cite{FuBerg} designed
to model the symmetries and parameters of Cu doped $Bi_{2}Se_{3}$. The case
that can be directly compared is when the relativistic mass term (denoted by
$m$ in ref.\cite{FuBerg}) is small compared to chemical potential. In this
work a more general effective electron - electron interaction was considered
with two couplings $V$ and $U$ for local intraband and interband
attractions, respectively. They are related to our $g$ by $g=2U=2V$.
Qualitatively, for $U/V=1$ one gets nearly degenerate energies (critical
temperatures were compared in ref.\cite{FuBerg}\ instead). This is similar
but not identical to our result without exchange, see Fig. 7. We indeed
obtain the degeneracy of the two gaps, the singlet and the triplet (their $%
\Delta _{1}$ and $\Delta _{2}$ respectively), but only in the limit of large
$g$. The gaps are definitely not degenerate when the coupling $g$ is below $%
20\pi ^{2}v_{F}^{3}\hbar ^{2}/T_{D}^{2}$. Even within the BCS regime (Eqs.(%
\ref{BCS_S},\ref{dT_BCS})), $\Delta _{T}/\Delta _{S}=\sinh \left(
0.35/\lambda \right) /\sinh \left( 0.5/\lambda \right) $. This is consistent
with $1$ only for quite large coupling (whatever UV cutoff used in ref. \cite%
{FuBerg}).

\bigskip

\textit{Acknowledgements.} We are indebted to C. W. Luo, J. J. Lin and W.B.
Jian for explaining details of experiments, and T. Maniv and M. Lewkowicz
for valuable discussions. Work of B.R. and D.L. was supported by NSC of
R.O.C. Grants No. 98-2112-M-009-014-MY3 and MOE ATU program. The work of
D.L. also is supported by National Natural Science Foundation of China (No.
11274018)

\end{document}